\documentclass{article}
\usepackage[margin=1.5in]{geometry}
\usepackage{amsmath}
\usepackage{amssymb}
\usepackage[T1]{fontenc}
\usepackage{url}
\usepackage[utf8]{inputenc} 
\usepackage{graphicx}
\usepackage{bm}
\usepackage{authblk}
\usepackage{longtable}
\usepackage{hyperref}
\usepackage{xcolor}

\title{Network localization strength regulates innovation diffusion with macro-level social influence}

\author[1]{Leyang Xue}
\author[2]{Kai-Cheng Yang}
\author[1]{Peng-Bi Cui\thanks{cuisir610@gmail.com}}
\author[1]{Zengru Di}
\affil[1]{International Academic Center of Complex Systems, Beijing Normal University, Zhuhai, 519087, China}
\affil[2]{Luddy School of Informatics, Computing, and Engineering, Indiana University, Bloomington, Indiana 47408, USA}

\date{}

\begin{document}
\maketitle

\newcommand{\SI}{Supplementary Information}

\begin{abstract}
Innovation diffusion in the networked population is an essential process that drives the progress of human society.
Despite the recent advances in network science, a fundamental understanding of network properties that regulate such processes is still lacking.
Focusing on an innovation diffusion model with pairwise transmission and macro-level social influence, i.e., more adopters in the networked population lead to a higher adoption tendency among the remaining individuals, we observe discontinuous phase transitions when the influence is sufficiently strong.
Through extensive analyses of a large corpus of empirical networks, we show that the tricritical point depends on the network localization strength, which our newly proposed metric can effectively quantify.
The metric reveals the deep connection between the critical and tricritical points and further indicates a trade-off: networks that allow less attractive products to prevail tend to yield slower diffusion and lower market penetration and verse versa.
Guided by this trade-off, we demonstrate how marketers can rewire the networks to modulate product diffusion according to their needs.
\end{abstract}



The dissemination of new ideas, technology, and products, or innovation diffusion, is a fundamental process that drives the socioeconomic development of our society.
Identifying the factors that affect such processes and analyzing their unique mechanism can yield invaluable insights and guide marketing strategies and policy-making~\cite{Rogers2003}.
Since it encompasses complex social processes such as interpersonal communication, social learning, decision-making, and social contagion, innovation diffusion has drawn strong research interest from various fields including management, economics, sociology, psychology, and physics~\cite{Chatterjee1990,Berry2018,Gabriel2014}. 
Due to the difficulty of excluding the effects of confounding factors in empirical studies, many researchers resort to the modeling approach~\cite{Bass1969,Kiesling2012}.

Naturally, the inherent characteristics of the technology or products and the psychology of the consumers can exert a strong impact on the diffusion process~\cite{Arts2011}.
But many recent empirical and simulation studies highlight the critical role played by the connectivity pattern of the target population~\cite{Jackson2010,Bale2013,Mccoy2014}. 
For instance, some structural features of networked population, such as heterogeneity and clustering, can largely shape the product dissemination process~\cite{Rahmandad2008,Peres2010}.
The hubs in the networks are of particular interest since they play a key role in promoting the product~\cite{Goldenberg2009,Delre2010}.
Such findings are not surprising since it is well-known in network science that underlying topological structures can strongly affect all dynamical processes~\cite{Castellano2009,Pastor2015}. 
However, a fundamental understanding of network properties that regulate the innovation diffusion process is still lacking and we aim to close this gap in this paper.

Following the recent trend in innovation diffusion modeling studies~\cite{Peres2010}, we consider the dissemination process of a new product in a networked population with social influence at different levels.
The micro-level influence is conveyed through pairwise interaction between neighbors in the network~\cite{Hohnisch2008} and represents mechanisms such as the well-documented word-of-mouth effect~\cite{Buttle1998}.
The macro-level influence increases the tendency of individuals to adopt the new product as market penetration increases.
Such influence is essentially a positive feedback mechanism between the macro system state and the micro pairwise transmission, which has many roots: With more adopters in the system, the remaining individuals are more likely to follow the social norm as a result of psychology effect~\cite{Mccoy2014,Young2009,Chen2016,Guilbeault2018} The utility of the product might increase  because of network externalities (typical examples are telecommunication products like fax, social networks, and smartphone)~\cite{Katz1992,Schoder2000,Rohlfs2003}; 
At the same time, the product price might decrease with the decline of marginal cost due to economies of scale~\cite{Kiesling2012,Scherer1990,Browning2020}.

We operationalize the diffusion process through a Susceptible-Infectious-Recovered (SIR)-like model~\cite{Pastor2015} where the transmissibility of the new product consists of two components.
The first part is a constant representing the inherent attractiveness of the product, while the second part is proportional to the number of adaptors in the population, representing the macro-level social influence.
By simulating this model on various empirical networks, we observe that the product can only occupy a viable market share when its transmissibility is larger than a critical value, analogous to the outbreak threshold of SIR model~\cite{Pastor2015}.
If the macro-level influence becomes strong enough, a shift of transition from continuous to discontinuous occurs. 
We use the tricritical point to refer to this threshold of macro-level influence.
We further find that products tend to achieve higher market penetration and diffuse faster on networks with smaller tricritical values and verse versa.

Due to their implications, we focus on the critical and tricritical points and find that the former can be approximated with the outbreak threshold of SIR model.
The latter, on the other hand, is closely related to the network localization strength~\cite{Goltsev2012,Travis2014}.
We propose a novel metric to measure such strength and demonstrate that our metric can accurately predict the tricritical point on a large corpus of empirical networks. 
The analytical expression of this metric automatically reveals its connection with the critical point and indicates a fundamental trade-off.
While holding the network density (i.e., number of edges) constant, a larger critical value of the transmissibility would yield a smaller tricritical point value of the macro-level influence and verse versa.
In other words, networks with stronger localization strength can facilitate the dissemination of less attractive products, yet lead to limited final market penetration. 
However, for networks with weaker localization strength, higher attractiveness is required for the product to spread out in an avalanche-like fashion, but the final market share tends to be larger. 
Guided by these findings, marketers can rewire the networks to fulfill their specific needs.


\section*{Model definition and behavior}
\label{subsec:model}

Consider the scenario where a new product is released to a networked population with $N$ individuals (nodes) and $L$ connections (edges).
Initially, most nodes are in the susceptible state (denoted by $S$) and may become adopted (denoted by $A$) after purchasing the product.
The adopters may introduce the product to their susceptible neighbors and successfully convert them with a probability $\beta^{'}$ through social influence.
Furthermore, with a probability $\mu$, the adopters lose their interest in advertising the new product and become recovered (denoted by $R$).
The dynamical process can be formulated as:
\begin{eqnarray}\label{eq:eqs1}
	S_i(t-1) + A_j(t-1) & \xrightarrow{\beta^{'}} & A_i(t) +A_j(t),\\
	A_i(t-1) & \xrightarrow{\mu} & R_i(t),
\end{eqnarray}
where $Y_i(t)$ stands for node $i$ in $Y\in \{S,A,R\}$ state at time $t$.
The transmission occurs only if there exists a connection between node $i$ and $j$.
The diffusion process stops when all adopters vanish.

\begin{figure}[t]
    \centering
    \includegraphics[width=\columnwidth]{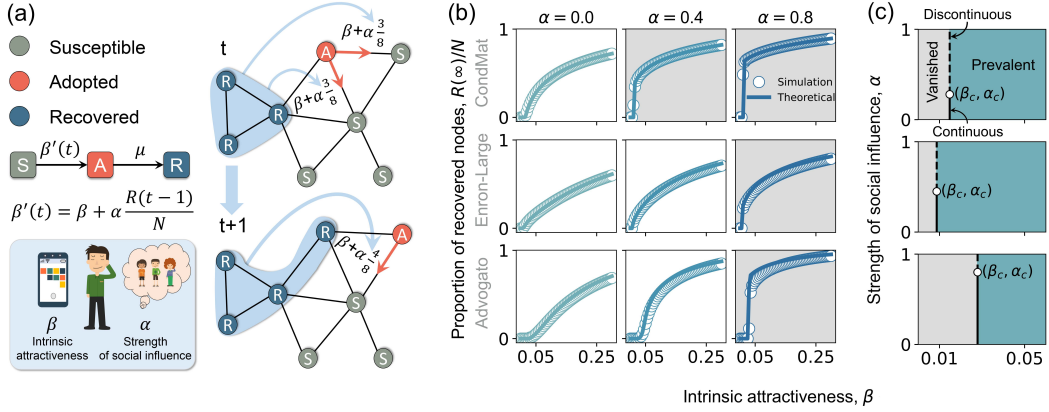}
    \caption{
    \textbf{Definition and behavior of the innovation diffusion model}. 
    (a) Schematic diagram of the model and its spreading process on a toy network with eight nodes. 
    The transmissibility $\beta^{'}(t)$ of the SIR-like model is determined by the intrinsic attractiveness $\beta$ of the new product and macro-level influence $\alpha\frac{R(t-1)}{N}$ together.
    At time $t$, the node in the adopted state is attempting to introduce the product to its two susceptible neighbors, and the contribution of macro-level influence to the transmissibility is $\alpha \frac{3}{8}$ since there are three recovered nodes in the system.
    At $t+1$, the intensity of macro-level influence becomes $\alpha \frac{4}{8}$ with one more recovered node added.
    (b) The market penetration, i.e., the final proportion of recovered nodes $R(\infty)/N$, for given $\alpha$ values as functions of $\beta$ on three real-world networks (Advogato, Enron-Large, and CondMat networks).
    The dots and the solid lines represent the simulation and theoretical results obtained by the dynamic message passing (DMP) method. 
    As the strength of macro-level influence increases, the diffusion processes start to exhibit discontinuous phase transitions (indicated by gray backgrounds).
    See the main text for details.
    (c) Phase diagrams for the same networks as (b).
    The diagrams are divided by the vertical line crossing the critical point $\beta_c$ into the vanished and prevalent states.
    The product fails to secure a viable market share in the former state but becomes very popular in the latter state.
    The solid~($\alpha<\alpha_c$) and dashed lines~($\alpha>\alpha_c$) denote continuous (CT) and discontinuous transitions (DT), respectively, and intersect at the tricritical point indicated by the white dot.}
    \label{fig:1}
\end{figure}

To incorporate both micro- and macro-level social influence documented in the literature~\cite{Bale2013,Mccoy2014,Delre2010}, we assume that the transmissibility $\beta^{'}$ follows the expression below:
\begin{eqnarray}
	\label{eq:eqs3}
	\beta^{'}(t) =  \min \left(1,\beta + \alpha \frac{R(t-1)}{N}\right), \alpha \geq 0, \beta^{'} \in \left[ 0,1\right]. 
\end{eqnarray}
$\beta$ represents the intrinsic attractiveness of the new product~\cite{Rogers2003}; $\alpha\frac{R(t-1)}{N}$ quantifies the macro-level influence, indicating that susceptible individuals are more likely to accept the product with more recovered ones in the population.
$R(t-1)$ is the number of recovered individuals in the system at time $t-1$ and $\alpha$ controls the strength of the macro-level influence.
$\beta^{'}(t)$ is further capped at one since it represents a probability.
The whole model is illustrated in Figure~\ref{fig:1}(a).

To characterize the model behavior, we simulate it with different values of $\beta$ and $\alpha$ on various empirical networks.
The results on three networks are shown in Figure~\ref{fig:1}(b).
We further adopt the dynamic message passing (DMP) method to theoretically track the final market penetration (i.e., $\frac{R(\infty)}{N}$). 
A good agreement between the theoretical predictions and simulations is illustrated in the figure.
More details of the simulation rules and the DMP method are provided in the Methods section.
Besides, a description of our network collection and supplementary results can be found in \SI{}.

Through extensive analyses, we identify several interesting phenomena.
In the absence of macro-level influence, i.e., $\alpha=0$, our diffusion model equates to the classical SIR model.
As $\beta$ increases above a critical value $\beta_c$, the system undergoes a continuous transition from the ``vanished'' state, where the new product fails to spread out, to the ``prevalent'' state, where a viable proportion of individuals finally adopts the product.
When the macro-level social influence is strong enough, discontinuous transitions start to emerge, i.e., the market penetration level experiences a sudden jump as $\beta$ increases~\cite{Lechman2018}.
In the meantime, the value of $\beta_c$ remains stable despite the changes in $\alpha$ (see \SI{} for further analyses). 
This is reasonable since the macro-level influence can only manifest when there are adoptors in the system, which requires $\beta\geq \beta_{c}$.

As expected, there is a critical value $\alpha_c$ above which the phase transition becomes discontinuous (see the plots with gray backgrounds in Figure~\ref{fig:1}(b))~\cite{Boccaletti2016,D2019}.
We define $\alpha_c$ as the tricritical point, whose value for a specific network can be numerically estimated (see the Methods section for details).
The model behavior can thus be characterized by the tuple $(\beta_c, \alpha_c)$ that splits the phase space into different regions, as shown in Figure~\ref{fig:1}(c). 
Like $\beta_c$, which determines whether a product with a given intrinsic attractiveness $\beta$ can successfully occupy the market, $\alpha_c$ also has real-world implications.
Specifically, networks with smaller $\alpha_c$ values tend to facilitate higher market penetration and faster product diffusion (see analyses in \SI{}).

\section*{Quantifying the critical and tricritical points of phase transition}

Due to the importance of $\beta_c$ and $\alpha_c$, estimating their values becomes crucial when studying innovation diffusion on networked population.
Although it is possible to obtain their values numerically, running large amounts of repeated simulations can be computationally heavy (see the Methods section for details).
Here, we show that the values of both $\beta_c$ and $\alpha_c$ are determined by the network structure and have simple algebra expressions based on the so-called non-backtracking matrix $\bm{B}$~\cite{Travis2014}.

\begin{figure}
    \centering
    \includegraphics[width=1.0\columnwidth]{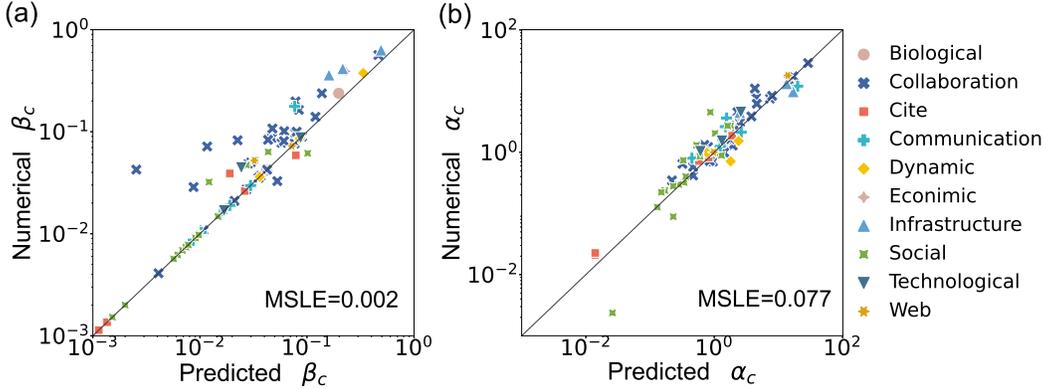}
    \caption{
    \textbf{Critical~($\beta_c$) and tricritical~($\alpha_c$) points for real-world networks}. 
    (a) Numerical values of $\beta_c$ obtained through simulations versus the predicted values, i.e., equation~(\ref{eq:eqs23}).
    (b) Numerical values of $\alpha_c$ obtained through simulations versus the predicted values generated by equation~(\ref{eq:alpha_l}).
    Each dot denotes a real-world network; solid diagonal lines of $y=x$ indicate an equality relationship between the predictions and the simulations.
    The mean squared logarithmic errors (MSLE) between the simulations and predictions are annotated in the plots.
    }
    \label{fig:2}
\end{figure}

In the previous section, we have established that the value of $\beta_c$ is independent of the macro-level influence.
Therefore, it can be approximated by the epidemic threshold of SIR model~\cite{Koher2019}:
\begin{eqnarray}\label{eq:eqs23}
	\beta_c = \frac{\mu}{\rho(\bm{B})+\mu-1},
\end{eqnarray}
where $\rho(\bm{B})$ denotes the largest eigenvalue of $\bm{B}$.
For simplicity, we set $\mu$ to 1, so our predicted value for the threshold is $1/\rho(\bm{B})$.
We compare the predicted and numerical values of the thresholds on 74 real-world networks in Figure~\ref{fig:2}(a).
The results are mostly in agreement with each other, along with several exceptions reported before~\cite{Pastor2020}.

The case for $\alpha_c$ is more complicated.
Compared to a continuous phase transition, much more susceptible individuals are converted to adopters simultaneously once $\beta$ exceeds the threshold during a discontinuous phase transition, finally leading to an avalanche-like growth of recovered population.
For this to happen, a positive feedback loop between more adopters (who will become recovered later) and larger transmissibility $\beta^{'}$ has to be established.
The presence of macro-level influence alone is insufficient; considerable individuals in the system need to share similar probabilities of becoming adopters as well.
While such probabilities typically vary from one individual to another since their connectivity patterns to others are different in most real-world networks~\cite{Pastor2015,Cai2015,Cui2018}.
For instance, a well-connected individual is more likely to be exposed to the new product and subsequently become an adopter than an individual with only a few neighbors in the network.
Previous studies suggest that this probability can be quantified by the non-backtracking centrality~\cite{Shrestha2015,Timar2022}, which is defined as
\begin{eqnarray}\label{eq:nbc}
    x_i = \sum_j A_{ij} v_{j\rightarrow i},
\end{eqnarray}
for node $i$.
$A_{ij}$ and $v_{j\rightarrow i}$ are the element of adjacency matrix \bm{$A$} and the component of principal eigenvector associated with the non-backtracking matrix \bm{$B$}, respectively.

\begin{figure}
  \centering
    \includegraphics[width=1.0\columnwidth]{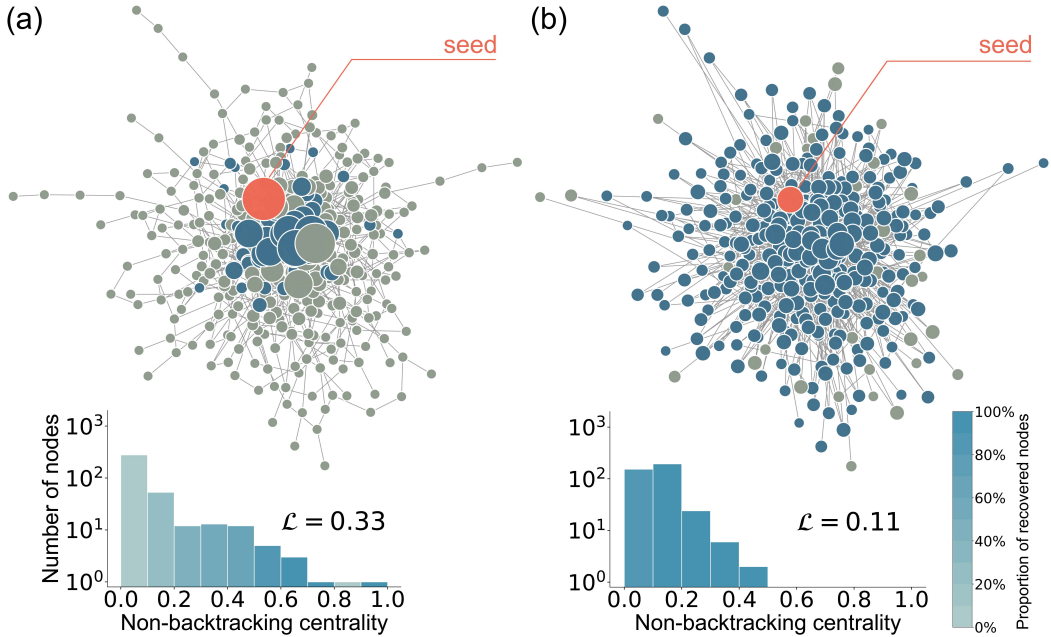}
    \caption{
    \textbf{Effect of network localization on innovation diffusion}.
    We simulate the product diffusion using the parameters $\alpha=1$ and $\beta=\beta_c$ (the theoretical critical value) on two synthetic networks and visualize the outcomes in (a) and (b), respectively.
    The network in (b) is obtained by randomly rewiring the edges of the real-world network in (a) while preserving the degree sequence.
    The size of each node corresponds to its non-backtracking centrality.
    Gray (blue) colors represent the susceptible (recovered) final states; the initial seed is highlighted in red.
    The histograms show the distribution of node non-backtracking centrality, and the color of each bar indicates the proportion of recovered nodes in the corresponding bin.
    We annotate the values of $\mathcal{L}$ for both networks in the figure.
    In terms of critical and tricritical values, we have $\beta_c=0.104, \alpha_c=1.17$ for (a) and $\beta_c =0.183, \alpha_c = 0.58$ for (b).
    }
    \label{fig:3}
\end{figure}

Therefore, we hypothesize that it is easier for discontinuous phase transitions to take place on networks with smaller variance among their node centrality measures.
To illustrate, we show the diffusion outcomes on two synthetic networks in Figure~\ref{fig:3}.
Two networks have the same degree sequence, but the discrepancy in their connectivity patterns leads to different non-backtracking centrality distributions, tricritical points, and final states.
Specifically, the network in Figure~\ref{fig:3}(a) has a small portion of nodes with much higher centrality indexes than other nodes, whereas the rewired network in Figure~\ref{fig:3}(b) has a more homogeneous centrality distribution.
By definition, the network in (a) has a stronger localization strength than that in (b).
As a result, the recovered nodes in (a) tend to have larger centrality indexes, i.e., the diffusion is constrained to the central nodes, while the recovered nodes are scattered across the whole network in (b).

Based on this observation, we define the localization strength $\mathcal{L}$ as:
\begin{eqnarray}\label{eq:localization_strength}
    \mathcal{L} = \frac{\sqrt{\frac{1}{N}\sum_{i=1}^{N}(x_i-\langle x_i \rangle)^2}}{\langle x_i \rangle \langle k \rangle}, 
\end{eqnarray}
where the numerator is the standard deviation of node non-backtracking centrality measures.
The denominator, i.e., the product of the average centrality value and the average degree, is added to ensure $\mathcal{L}$ is comparable across different networks.
We annotate the $\mathcal{L}$ values for both networks in Figure~\ref{fig:3} to illustrate that this metric can quantify the network localization strength.

By analyzing $\mathcal{L}$ and $\alpha_c$ for different networks, we find a positive correlation between them.
So we use the following formula to capture this relationship:
\begin{eqnarray}\label{eq:alpha_l}
   \alpha_c = \lambda \mathcal{L}^\eta,
\end{eqnarray}
where $\lambda$ and $\eta$ need to be determined empirically.
To avoid overfitting on the real-world networks, we fit equation~(\ref{eq:alpha_l}) on a collection of synthetic networks and obtain the results $\lambda = 2.63, \eta=1.00$ (see the Methods section for details).
We then compare the simulated values of $\alpha_c$ with their predicted values, i.e., $2.63 \mathcal{L}$, for our collection of real-world networks in Figure~\ref{fig:2}(b) and find an excellent agreement.
Using this relationship, one can easily predict the $\alpha_c$ value for any network as long as its topology is known, without the need to run extensive simulations.
In practice, one can fit equation~(\ref{eq:alpha_l}) directly on the real-world networks at hand for better estimates of $\lambda$ and $\eta$, 

We note that there are other methods to quantify network localization strength. 
For instance, the inverse participation ratio (IPR) has been widely used in the literature for this purpose~\cite{Travis2014}.
One can also replace $x_i$ in equation~(\ref{eq:localization_strength}) with eigenvector centrality~\cite{Bonacich1972}, or use Gini coefficient to measure the variance~\cite{Bendel1989}.
However, the definition given by equation~(\ref{eq:localization_strength}) provides the best predictions for $\alpha_c$ (see \SI{} for comparison).

\section*{Relationship between the critical and tricritical points}

\begin{figure}
    \centering
    \includegraphics[width=1.0\columnwidth]{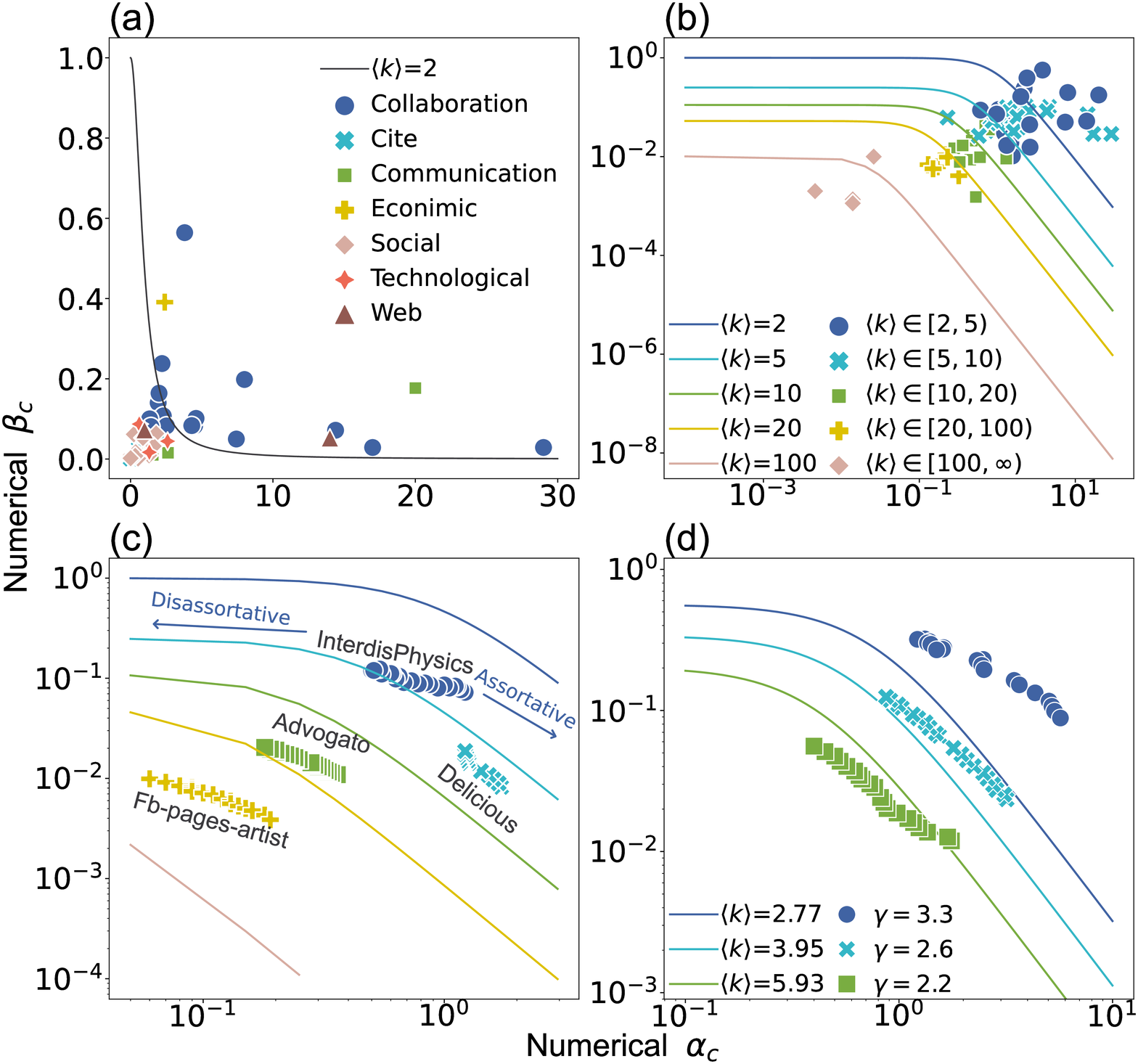}
    \caption{\textbf{Relationship between the critical and tricritical points}.
    $\beta_{c}$ and $\alpha_{c}$ are obtained though numerical simulations.
    Each marker corresponds to a network, and its location indicates the corresponding critical and tricritical values.
    The solid lines represent the relationship between $\beta_c$ and $\alpha_c$ described by equation~\ref{eq:beta_alpha_relation} with given mean degree.
    (a) We plot equation~\ref{eq:beta_alpha_relation} with $\langle k \rangle = 2$.
    (b) We mark the networks according to their average degree and show the lines corresponding to different $\langle k \rangle$ values.
    The plot is in log-scale to highlight the details.
    (c) By changing network assortativeness, we obtain configurations with different $\beta_c$ and $\alpha_c$ values for four selected real-world networks~(i.e., InterdisPhysics, Delicious, Advogato, Fb-pages-artist).
    The directions of the changes are annotated in the plot.
    (d) Same as (c) but with three instances of scale-free networks generated by the configuration model.
    }
    \label{fig:4}
\end{figure}

In addition to predicting the tricritical point, Equation~(\ref{eq:alpha_l}) allows us to explore the relationship between $\beta_c$ and $\alpha_c$ analytically.
Specifically, on random uncorrelated networks, we have
\begin{eqnarray}\label{eq:beta_alpha_relation}
   \beta_c = \frac{1}{\langle k\rangle^3 (\alpha_c/\lambda)^{2/\eta} +\langle k \rangle -1},
\end{eqnarray}
where average degree $\langle k\rangle$ quantifies the edge density of the network.
See the Methods section for the derivation of equation~(\ref{eq:beta_alpha_relation}).
Given all the networks considered have $\langle k\rangle > 2$, equation~\ref{eq:beta_alpha_relation} suggests that $\beta_c$ decreases monotonically with $\alpha_c$ and larger $\langle k \rangle$ can lead to smaller values for both $\beta_c$ and $\alpha_c$.

We mention earlier that a small $\beta_c$ means that a less attractive product can still occupy the market while a small $\alpha_c$ indicates faster diffusion and higher market penetration.
Therefore, the extent to which a real-world network can popularize the innovation i.e. its efficiency is negatively correlated with the values of $\beta_c$ and $\alpha_c$. 
To qualitatively give a judgement of the efficiencies of real-world networks, in Figure~\ref{fig:4}(a), we plot a boundary through equation~(\ref{eq:beta_alpha_relation}) with the limiting value $\langle k \rangle=2$, together with markers of $\beta_c$ and $\alpha_c$ for the collection of real-world networks.
The boundary gives the performance of the sparsest tree-like network, to a great extent, and can thus be used to judge whether or not a network is effective.
We observe that most networks reside below the boundary, which is also located close to the original point, and obviously effective.
There are a handful of inefficient exceptions belonging to classes of collaboration, communication, economics, and Web.
This is especially so for most collaboration networks, showing further distance to the origin of coordinates, given by relatively large $\alpha_c$ and small $\beta_c$.
This pattern might be due to the negative effects of cultural or knowledge barriers in knowledge transfer~\cite{Sheng2013}. It is evident that the efficiency of a real-world network can be roughly estimated by the inverse of its distance to the origin of the coordinate in plane of $(\beta_{c}, \alpha_{c})$.  

We then reproduce Figure~\ref{fig:4}(a) in (b) using logarithmic scale to better observe the details.
This time, we show the troughs for equation~(\ref{eq:beta_alpha_relation}) with multiple $\langle k \rangle$ values, and color the networks based on the average degree brackets to which they belong.
We find that most networks are located within the troughs predicted by equation~(\ref{eq:beta_alpha_relation}), confirming our theory about the relationship between the critical and tricritical points.

Besides, this relationship has important implications.
For the best diffusion outcome, one wishes $\beta_c$ and $\alpha_c$ to be small simultaneously.
According to equation~(\ref{eq:beta_alpha_relation}), this can only happen on dense networks.
It can also be concluded from the results reported in Figure~\ref{fig:4} that the efficiency of a network, claimed to have  negative correlation with the values of $\beta_c$ and $\alpha_c$, is  positively associated with network density, but negatively associated with localization strength $\mathcal{L}$.  
Consequently, when the network density is fixed, equation~(\ref{eq:beta_alpha_relation}) imposes a trade-off in popularizing the products.
That is, strong localization (high $\mathcal{L}$) which causes smaller $\beta_c$ can inevitably lead to larger $\alpha_c$, and vice versa.
Interestingly, the results in Figure~\ref{fig:4} suggest that most empirical networks considered here strike a balance between $\beta_c$ and $\alpha_c$.

To reduce the value of $\beta_c$ and $\alpha_c$, one can consider adding edges to the network to increase its density.
But when this is not feasible, we find that rewiring the network while preserving the edge density can also change the values of $\beta_c$ and $\alpha_c$.
Here we demonstrate one effective approach, the Xalvi-Brunet $\&$ Sokolov algorithm~\cite{Xulvi2004}, to achieve this goal.
It was originally proposed to tune the assortativeness of the networks, but we find that it can also modify the localization strength~(see the \SI{} for details).
Specifically, increasing the assortativeness of the network can enhance its localization strength, further yielding larger $\alpha_c$ in exchange for smaller $\beta_c$.
We show the results on four selected networks in Figure~\ref{fig:4}(c) and find that the constraint from equation~(\ref{eq:beta_alpha_relation}) is still in effect during this process.
The technique also works on synthetic networks, see examples in Figure.~\ref{fig:4}(d).

\section*{Discussion and outlook}

In this paper, we study the innovation diffusion process on networked populations and reveal the fundamental topological structure that regulates this process.
We focus on a model that encapsulates the micro-level pairwise transmission and macro-level influence observed in empirical studies.
In the absence of the macro-level influence, the model exhibits a continuous phase transition from the vanished state to the prevalent state as we increase product attractiveness.
The transition becomes discontinuous when the influence is strong enough.
One main contribution is the finding that network localization strength regulates the innovation diffusion process by determining the critical and tricritical values.
We provide a novel analytical expression for localization strength, making it possible to estimate the tricritical point purely based on network topology.
It advances a long line of modeling efforts to understand the impact of network structure on innovation diffusion from multiple research fields, including marketing, economics, and physics. 

The critical and tricritical points have real-world implications: the former determines whether the product can occupy the market, and the latter determines the speed of diffusion and the final adoption rate.
Ideally, one wishes to have small critical and tricritical values simultaneously.
Our findings suggest that this can be achieved by adding more edges to the network.
But when the edge density of the network is held constant, the relationship between the critical and tricritical points indicates that reducing one would increase the other.

Marketers can use our framework to create tailored marketing strategies.
For products with high intrinsic attractiveness, they can choose a target population with relatively larger $\beta_c$.
This could lead to a rapid occupation of the market with higher penetration.
On the other hand, if the intrinsic attractiveness of the product is limited, then the ideal target population should have a smaller $\beta_c$.
Even though the diffusion process is not as rapid and the final adoption rate is not as high, the product can at least secure a viable market share.
Alternatively, one can also consider modifying the network structure for a fixed target population.
For instance, adding new connections, i.e., channels of diffusion, to the network can effectively facilitate product diffusion.
One can also adjust the connectivity patterns and trade one metric for the other to optimize the network for a specific product.

Our work leads to many exciting new research questions and lays the foundation for further exploration.
For instance, we apply many assumptions and approximations while deriving the relationship between the critical and tricritical values, including treating all networks as random ones without degree correlation.
This obviously fails to capture the features of real-world networks and leads to the discrepancy between the predicted and numerical results.
How to accurately characterize such a relationship remains an open question.
Another important direction is to develop cost-efficient network modification approaches to adjust the localization strength as needed.
For instance, identifying a minimum set of edges that need to be added or rewired to achieve the goal is of great practical value.

Our findings also have potential applications beyond the innovation diffusion context.
The localization phenomena play an essential role in many dynamical processes occurring on complex systems, especially epidemic spreading~\cite{Goltsev2012,Pastor2020}.
Here we show that the newly proposed metric for quantifying localization strength works better than previous ones in predicting the position of the tricritical point.
Moreover, its relatively simple algebra expression makes analytical treatment possible.
It would be of particular interest to see what new insights it can bring when applied to different contexts.

\section*{Methods}

\subsection*{Numerical simulations}

In each realization, a node is randomly assigned the $A$ state to initiate the diffusion, while the remaining nodes are in the $S$ state.
The diffusion process proceeds following the model rules until the absorbing state is reached, where no transmission can occur anymore.
Throughout this paper, the recovery probability $\mu$ is set to $1$.

In our model, the position of critical point $\beta_c$ is numerically identified by searching for the maximum value of the susceptibility
\begin{equation}\label{eq:chi}
\chi =\frac{\sqrt{\langle r(\infty)^2\rangle  - \langle r(\infty) \rangle^2}}{\langle r(\infty) \rangle}.
\end{equation}
For reliability, we perform $10^4$ independent realizations to calculate the first-order and second-order moments of $r(\infty)$ on a given network with the same parameters.

To identify the value of $\alpha_c$, we categorize the phase transitions into different classes by plotting the mass distribution of $r(\infty)$ near the predicted $\beta_c$ values, with varying $\alpha$ for a large number of independent realizations.
Discontinuous phase transitions occur if and only if the isolated peak of giant recovered clusters emerges~\cite{Cai2015}.
Details and examples can be found in \SI{}.

\subsection*{Non-backtracking matrix and non-backtracking centrality}

Since most of our theoretical analyses depend on non-backtracking matrix, here we provide a brief introduction to it.
Considering an undirected network with $L$ edges, its non-backtracking matrix $\bm{B}$ is a $2L \times 2L$ non-symmetric matrix whose rows and columns denote directed edges $j \rightarrow i$ pointing from node $j$ to $i$. 
And we have:
\begin{eqnarray}\label{eq:eqs19}
	B_{z \rightarrow i,j\rightarrow z'} = \left \{
	\begin{aligned}
		1   \qquad    &  if \quad z' = z, j \neq i,\\
		0   \qquad     & otherwise. \\
	\end{aligned}
	\right.
\end{eqnarray}

When all nodes in the network belong to a strongly connected giant component, the non-backtracking matrix is non-negative and irreducible.
According to the Perron-Frobenius theorem~\cite{Horn2012}, there exists a positive leading eigenvector $v_{j \rightarrow i}$ associated to the largest eigenvalue $\lambda_B$:
\begin{equation}
    \lambda_B v_{j \rightarrow i} = \sum_{k\rightarrow m} B_{j\rightarrow i,k\rightarrow m} v_{k\rightarrow m}.
\end{equation}
The element $v_{j \rightarrow i}$ of the leading vector represents the centrality of node $j$ while ignoring any contribution from the node $i$. 
The non-backtracking centrality of node $i$ is defined as 
\begin{equation}\label{eq:eqs25}
    x_i = \sum_{j}A_{ij}v_{j \rightarrow i}.
\end{equation}
Unlike eigenvector centrality, non-backtracking centrality can reduce the effect of the ``self-inflating'' phenomenon associated with the hub nodes~\cite{Travis2014,Pastor2020}.

\subsection*{Dynamic Message-Passing method}
\label{subsubsec:dmp}

We use the dynamic message-passing method (DMP) to analytically treat the outbreak size and threshold.
DMP has been used extensively in the context of contagion processes.
It can prevent the contagion from backtracking to the source node, avoiding the mutual transmission effect~\cite{Karrer2010,Lokhov2014,Koher2019}.
Regardless of the initial conditions, DMP can accurately predict the probability of each node being in a specific state at time $t$, especially for tree-like networks.

First, let us derive the exact equations of DMP.
The probabilities of node $i$ being in $S$, $A$, or $R$ states at time $t$ are represented by $P^i_{S}(t)$, $P^i_{A}(t)$ or, $P^i_{R}(t)$, respectively. The three terms follow the constraint:
\begin{equation}\label{eq:sum_to_one}
    P_A^i(t) + P_S^i(t) + P_R^i(t) = 1.
\end{equation}

The recovery process for an adopted node $i$ is independent of its neighbors' state, and $P^i_{R}(t)$ can thus be represented as
\begin{equation}\label{eq:eqs4}
	P_R^i(t) = P_R^i(t-1) + \mu P_A^i(t-1).
\end{equation}
Subsequently, the global quantity $R(t)$ can be obtained using
\begin{eqnarray}\label{eq:eqs5}
	R(t) = \sum_i^N{P_R^i(t)}.
\end{eqnarray}
Substituting equation~(\ref{eq:eqs5}) into equation~(\ref{eq:eqs3}), we have
\begin{eqnarray}\label{eq:eqs7}
	\beta^{'}(t) = \min\left(1,\beta + \alpha\frac{\sum^{N}_{i=1}P^{i}_{R}(t-1)}{N}\right),
\end{eqnarray}
which increases with $R(t-1)$ until it reaches the upper bound $\beta^{'}(t) = 1$.

Next, $P^i_{S}(t)$ can be expressed as 
\begin{eqnarray}\label{eq:eqs8}
	P_S^i(t) =  P_S^i(0) \Phi_i(t), 
\end{eqnarray}
where $\Phi_i(t)$ denotes the probability that the product has not been successfully transmitted to node $i$ from its adopted neighbors until time $t$.
DMP assumes that the underlying network is tree-like and the state evolution of node $i$'s neighbors is independent.
However, once node $i$ is successfully persuaded by its neighbor $z$ and switches to $A$ state, it would attempt to convince another neighbor $z'$ ($z' \neq z$).
It gives rise to a dilemma because the state transitions of node $z'$ and $z$ are clearly correlated.
To alleviate this issue, we assume that the focal node $i$ is a cavity and define $\theta^{z\rightarrow i}(t)$ as the probability that node $i$ has not been successfully persuaded by node $z$ until time $t$.
$\Phi_i(t)$ can then be factorized as $\prod_{z\in \partial i} \theta^{z\rightarrow i}(t)$ where $\partial i$ is the neighbor set of node $i$.
Substituting it into equation~(\ref{eq:eqs8}), we have
\begin{eqnarray}\label{eq:eqs9}
	P_S^i(t) = P_S^i(0) \prod_{z\in \partial i} \theta^{z\rightarrow i}(t). 
\end{eqnarray}

Message-passing is directional, meaning that $\theta^{i\rightarrow z}(t) \neq \theta^{z\rightarrow i}(t)$ for undirected networks.
In our analysis, we treat each undirected edge as two directed edges pointing to opposite directions.
Initially, we have $\theta^{z \rightarrow i}(0) = 1$ for all edges in the network.
Later, $\theta^{z \rightarrow i}(t-1)$ decreases as the product transmits from node $z$ to $i$, which occurs with the probability $\beta^{'}(t)\phi^{z\rightarrow i}(t-1)$ where $\phi^{z\rightarrow i}(t-1)$ is the probability that the adjacent adopted node $z$ has not passed the product to node $i$ until time $t-1$.
Hence, $\theta^{z \rightarrow i}(t)$ follows the updating rule:
\begin{eqnarray}\label{eq:eqs10}
	\theta^{z\rightarrow i}(t) = \theta^{z\rightarrow i}(t-1) - \beta^{'}(t) \phi^{z\rightarrow i}(t-1).
\end{eqnarray}

Now we need to find the expression for $\phi^{z \rightarrow i}(t)$.
On the one hand, $\phi^{z \rightarrow i}(t)$ decreases when node $z$ in $A$ state recovers, or when the product is successfully transmitted from $z$ to $i$, or when the two processes occur simultaneously.
The corresponding probabilities for these events are $\mu$, $\beta^{'}(t)$, and $\mu \beta^{'}(t)$, respectively.
On the other hand, $\phi^{z \rightarrow i}(t)$ increases when node $z$ in $S$ state becomes $A$.
The changing rate $\Delta P^{z\rightarrow i}_S(t)$ can be calculated as $P_S^{z\rightarrow i}(t-1) - P_S^{z\rightarrow i}(t)$ where $P_S^{z\rightarrow i}(t)$ represents the probability of node $z$ staying in $S$ state after interacting with the cavity node $i$.
By combining all terms we have
\begin{eqnarray}\label{eq:eqs11}
		\phi^{z\rightarrow i}(t) & = & \phi^{z\rightarrow i}(t-1) - \beta^{'}(t)\phi^{z\rightarrow i}(t-1) -\mu\phi^{z\rightarrow i}(t-1) + \mu\beta^{'}(t)\phi^{z\rightarrow i}(t-1) + P_S^{z\rightarrow i}(t-1) - P_S^{z\rightarrow i}(t)\notag \\
		& = & (1-\beta^{'}(t))(1-\mu)\phi^{z\rightarrow i}(t-1) + P_S^{z\rightarrow i}(t-1) - P_S^{z\rightarrow i}(t). 
\end{eqnarray}

The next step is to express $P_S^{z \rightarrow i}(t)$ explicitly.
Since node $i$ is a cavity, node $z$ will stay susceptible as long as it is not transfered by another neighbor $j$.
Using equation~(\ref{eq:eqs9}), we can obtain the probability that $z$ remains susceptible when its neighbor $i$ is a cavity: 
\begin{eqnarray}\label{eq:eqs12}
	P_S^{z \rightarrow i}(t) = P_S^z(0) \prod_{j\in\partial z\backslash i} \theta^{j\rightarrow z}(t),
\end{eqnarray}
where $\partial z\backslash i$ represents the neighbors of $z$ except for $i$.
At $t=0$, we have $P_S^{z\rightarrow i}(0)=0$ if $z$ is the seed that initiates the diffusion process and $P_S^{z\rightarrow i}(0)=1$ otherwise.
This can be represented as $P_S^{z\rightarrow i}(0) = 1-\delta_{q_z(0),A}$ where $\delta_{q_z(0),A}$ is the Kronecker function and $q_z(0)$ represents the state of $z$ at time $t=0$.

Finally, we can employ equations~(\ref{eq:eqs7}), (\ref{eq:eqs10})-(\ref{eq:eqs12}) to track the exact trajectory of $\theta^{z \rightarrow i}(t)$, $\phi^{z \rightarrow i}(t)$, and $P_S^{z \rightarrow i}(t)$ with the following initial conditions:
\begin{eqnarray}\label{eq:eqs13}
\theta^{z\rightarrow i}(0) & = &  1, \\
\phi^{z\rightarrow i}(0) & = & P^z_A(0) = \delta_{q_z(0),A},\\
P^{z\rightarrow i}_S(0) & = & P^z_S(0) =  1-\delta_{q_z(0),A}. 
\end{eqnarray}
By further combining $P_R^i(0)=0$ and $\beta(0)=\beta$ with equations~(\ref{eq:sum_to_one}), (\ref{eq:eqs4}), (\ref{eq:eqs7}), and (\ref{eq:eqs9})-(\ref{eq:eqs12}) we can calculate the trajectory of $P^i_{S}(t)$, $P^i_{A}(t)$, $P^i_{R}(t)$, and obtain the order parameter $R(\infty)$.
The computation complexity of DMP is $O(L)$, where $L$ is the number of edges in the network.

\subsection*{Estimating the fitting parameters of localization strength}

To characterize the relationship between localization strength $\mathcal{L}$ and $\alpha_c$, we first calculate their values for various network models and real-world networks (see \SI{} for details) and plot them. 
We find that one grows linearly with the other in the log-scale plots so we assume they follow the relationship $ln \alpha_c = \eta ln \mathcal{L} + \delta$, or $\alpha_c = \lambda \mathcal{L}^{\eta}, \lambda = e^{\delta}$.
The values of $\eta$ and $\delta$ need to be estimated from data.
To avoid overfitting the real-world networks, we use the least square method to fit the relationship on a group of network models and obtain $\eta = 1$ and $\lambda = 2.63$.
Therefore, we have $\alpha_c = 2.63 \mathcal{L}$.
This allows estimating the value of $\alpha_c$ purely based on the network topology.

\subsection*{Relationship between the critical and tricritical points}

With the analytical expressions of $\alpha_c$ and $\beta_c$, we can examine their relationship.
For uncorrelated random networks, the element on the primary eigenvector in the non-backtracking matrix depends on the degree of nodes, see the equation~(\ref{eq:eqs26})~\cite{Pastor2020},
\begin{eqnarray}\label{eq:eqs26}
    v_{j\rightarrow i} \sim k_j -1.
\end{eqnarray}
The non-backtracking centrality for node $i$ can then be approximated as
\begin{equation}\label{eq:eqs27}
    x_i = \sum_{j}A_{ij}v_{j \rightarrow i} \sim \sum_j{\frac{k_i k_j}{N \langle k \rangle}(k_j-1)} =\frac{\langle k^2 \rangle - \langle k \rangle}{\langle k \rangle} k_i,
\end{equation}
where we replace the $A_{ij}$ with its expression in annealed networks, namely $\hat{A}_{ij} = k_i k_j/N \langle k \rangle$.
The first-order moment of $x_i$ can be expressed as 
\begin{eqnarray}\label{eq:eqs28}
    \langle x_i \rangle = \frac{\sum_i x_i}{N} \sim \langle k^2 \rangle - \langle k \rangle.
\end{eqnarray}
By substituting equations~(\ref{eq:eqs27}) and (\ref{eq:eqs28}) into equation~(\ref{eq:localization_strength}), we have 
\begin{eqnarray}\label{eq:eqs29}
    \mathcal{L} = \frac{\sqrt{\langle k^2 \rangle - {\langle k \rangle}^2}}{{\langle k \rangle}^2}.
\end{eqnarray}
This suggests that $\mathcal{L}$ only depends on the first- and second-order moments of $\langle k \rangle$.
Similarly, $\beta_c$ can be represented as
\begin{eqnarray}\label{eq:eqs30}
    \beta_c = \frac{\langle k \rangle}{\langle k^2 \rangle - \langle k \rangle},
\end{eqnarray}
on annealed networks~\cite{Pastor2020}.

Combining equations~(\ref{eq:eqs29}) and (\ref{eq:eqs30}), we have 
\begin{eqnarray}\label{eq:eqs31}
   \beta_c = \frac{1}{\langle k\rangle^3 \mathcal{L}^2 +\langle k \rangle -1}.
\end{eqnarray}
And according to the relationship $\alpha_c = \lambda \mathcal{L}^\eta$, we obtain 
\begin{eqnarray}\label{eq:eqs32}
   \beta_c = \frac{1}{\langle k\rangle^3 (\alpha_c/\lambda)^{2/\eta} +\langle k \rangle -1}.
\end{eqnarray}

\section*{Data availability}
All data supporting this study are available on Mendeley Data~(\url{https://data.mendeley.com/datasets/d848h7rcdg}) and are described in the \SI{}. 

\section*{Code availability}
The code to run numerical simulations and to plot the figure is available on GitHub~(\url{https://github.com/LeyangXue/InnovationDiffusion}).

\section*{Acknowledgements}
This work was supported by the Key Program of the National Natural Science Foundation of China (Grant No.~71731002), by the National Natural Science Foundation of China (Grant Nos.~), and by the Guangdong Basic and Applied Basic Research Foundation (Grant No.~2021A1515011975). We thank Claudio Castellano for the helpful comments and suggestions.

\section*{Author contributions}
All authors have contributed equally to this article.
\section*{Competing interests}
The authors declare no competing interests.

\section*{Supplementary information: network localization strength regulates innovation diffusion with macro-level social influence}

\numberwithin{equation}{section}
\section{Networks}

We use various networks in this paper.
Below we describe their sources and characteristics.

\subsection{Real-world networks}

We compile a large collection of real-world networks that facilitate innovation diffusion. 
They are from the following channels:
(1) Network Repository\footnote{\url{https://networkrepository.com/}} - NetRes~\cite{Ryan2015};
(2) AMiner\footnote{\url{https://www.aminer.cn/data/}}~\cite{Cios2005};
(3) Pajek\footnote{\url{http://vlado.fmf.uni-lj.si/pub/networks/data/default.htm}}~\cite{Batagelj2006};
(4) American physical society\footnote{\url{https://journals.aps.org/datasets}} - APS;
(5) Index of Complex Network\footnote{\url{https://icon.colorado.edu/\#!/networks}} - ICN;
(6) Stanford Network Analysis Project\footnote{\url{http://snap.stanford.edu/data/}} - SNAP~\cite{Leskovec2014};
(7) Github\footnote{\url{https://github.com/benedekrozemberczki/datasets}};
(8) Kaggle\footnote{\url{https://www.kaggle.com/datasets/andrewlucci/huawei-social-network-data}};
(9) Network Science by Albert-Laszlo Barabasi\footnote{\url{http://networksciencebook.com/translations/en/resources/data.html}} - Barabasi~\cite{Barabasi2016};
These networks describe connectivity patterns of different systems, including collaboration networks, citation networks, communication networks, technological networks, and social networks.
We only keep the largest connected component from each network.
The details of these networks can be found in Supplementary Table~\ref{tab:Tab1}).

\setlength{\tabcolsep}{1.0pt}
\setlength{\LTcapwidth}{\textwidth}
{
\scriptsize
\begin{longtable}{lrrrrrrrrrrrrr}
     \caption{Properties of real-world networks.
     We report network name, network type, number of nodes $N$, number of edges $L$, average degree $\langle k \rangle$, degree exponent $\gamma$ estimated by the maximum likelihood method, degree correlation coefficient $r$, predicted critical point $\beta_c$, numerical critical point $\beta_c^{num}$, localization strength $\mathcal{L}$, predicted tricritical point $\alpha_c$, numerical tricritical point $\alpha_c^{num}$, and the source of networks.}
     \label{tab:Tab1}
     \\ \hline
     \multicolumn{1}{c}{Network} & \multicolumn{1}{c}{Type} & \multicolumn{1}{c}{$N$} & \multicolumn{1}{c}{$L$} & \multicolumn{1}{c}{$\langle k \rangle$} & \multicolumn{1}{c}{$\gamma$} & \multicolumn{1}{c}{$r$} & \multicolumn{1}{c}{$\beta_c$} & \multicolumn{1}{c}{$\beta_c^{num}$} & \multicolumn{1}{c}{$\mathcal{L}$} & \multicolumn{1}{c}{$\alpha_c$} & \multicolumn{1}{c}{$\alpha_c^{num}$} & \multicolumn{1}{c}{Source}\\ \hline
     \endfirsthead
     \\\hline
     \multicolumn{1}{c}{Network} & \multicolumn{1}{c}{Type} & \multicolumn{1}{c}{$N$} & \multicolumn{1}{c}{$L$} & \multicolumn{1}{c}{$\langle k \rangle$} & \multicolumn{1}{c}{$\gamma$} & \multicolumn{1}{c}{$r$} & \multicolumn{1}{c}{$\beta_c$} & \multicolumn{1}{c}{$\beta_c^{num}$} & \multicolumn{1}{c}{$\mathcal{L}$} & \multicolumn{1}{c}{$\alpha_c$} & \multicolumn{1}{c}{$\alpha_c^{num}$} & \multicolumn{1}{c}{Source} \\ \hline
     \endhead
     \hline 
     \multicolumn{3}{l}{{Continued on next page}} \\ 
     \endfoot
     \hline
     \endlastfoot
    bio-dmela & Biolog. & 7,393 & 25,569 & 6.92 & -5.06 & -0.05 & 0.0419 & 0.0419 & 0.2923 & 0.7395 & 0.5800 & NetRes \\ 
    bio-yeast-protein-inter & Biolog. & 1,458 & 1,948 & 2.67 & -2.98 & -0.21 & 0.1980 & 0.2380 & 0.9141 & 2.3126 & 1.7700 & NetRes \\
    ca-aminer & Collab. & 942,212 & 3,808,259 & 8.08 & -4.39 & 0.10 & 0.0088 & 0.0288 & 10.9531 & 27.7113 & 29.0000 & NetRes \\
    ca-BayesianNet & Collab. & 554 & 1,238 & 4.47 & -2.34 & 0.32 & 0.0471 & 0.0871 & 0.7152 & 1.8094 & 2.3400 & Aminer \\
    ca-citeseer & Collab. & 227,320 & 814,134 & 7.16 & -3.13 & 0.07 & 0.0117 & 0.0717 & 5.8610 & 14.8284 & 14.4000 & NetRes \\
    ca-CSphd & Collab. & 1,025 & 1,043 & 2.04 & -2.28 & -0.25 & 0.4642 & 0.5642 & 1.4680 & 3.7140 & 3.8000 & NetRes \\
    ca-Data-Mining & Collab. & 679 & 1,687 & 4.97 & -3.02 & 0.12 & 0.0767 & 0.0767 & 0.3837 & 0.9708 & 1.2600 & Aminer \\
    ca-Database-System & Collab. & 1,127 & 6,690 & 11.87 & -2.76 & 0.21 & 0.0271 & 0.0271 & 0.1587 & 0.4014 & 0.4700 & Aminer \\
    ca-dblp-2012 & Collab. & 317,080 & 1,049,866 & 6.62 & -3.26 & 0.27 & 0.0087 & 0.0287 & 6.6393 & 16.7974 & 17.0000 & NetRes \\
    ca-Information-Fusion & Collab. & 348 & 595 & 3.42 & -2.77 & 0.17 & 0.1379 & 0.2379 & 0.8439 & 2.1351 & 2.2100 & Aminer \\
    ca-Information-Retrieval & Collab. & 657 & 1,907 & 5.81 & -2.50 & 0.36 & 0.0424 & 0.0424 & 0.4814 & 1.2179 & 1.4400 & Aminer \\
    ca-Machine-Learning & Collab. & 920 & 2,285 & 4.97 & -3.68 & 0.09 & 0.0899 & 0.0899 & 0.3584 & 0.9068 & 1.0400 & Aminer \\
    ca-Semantic-Web & Collab. & 671 & 2,237 & 6.67 & -4.30 & 0.21 & 0.0528 & 0.0328 & 0.2687 & 0.6798 & 0.9500 & Aminer \\
    ca-Web-Services & Collab. & 400 & 777 & 3.89 & -3.46 & 0.14 & 0.1200 & 0.1400 & 0.4896 & 1.2387 & 1.9400 & Aminer \\
    ca-Erdos02 & Collab. & 5,534 & 8,472 & 3.06 & -1.93 & -0.04 & 0.0579 & 0.0779 & 0.6303 & 1.5947 & 2.0200 & NetRes \\
    ca-Erdos992 & Collab. & 4,991 & 7,428 & 2.98 & -8.90 & -0.45 & 0.0777 & 0.0977 & 1.0423 & 2.6369 & 1.7400 & NetRes \\
    ca-Math-compugeom & Collab. & 3,621 & 9,461 & 5.23 & -2.44 & 0.17 & 0.0369 & 0.0369 & 0.6281 & 1.5891 & 1.8000 & Pajek \\
    ca-MathSciNet & Collab. & 332,689 & 820,644 & 4.93 & -5.03 & 0.10 & 0.0298 & 0.0498 & 2.8769 & 7.2784 & 7.4300 & NetRes \\
    ca-AstroPh & Collab. & 17,903 & 196,972 & 22.00 & -4.50 & 0.20 & 0.0108 & 0.0108 & 0.1319 & 0.3337 & 0.2200 & NetRes \\
    ca-AstroPhysTechObs & Collab. & 3,237 & 10,569 & 6.53 & -6.01 & 0.29 & 0.0476 & 0.1076 & 1.7281 & 4.3721 & 2.3000 & APS \\
    ca-CondMat & Collab. & 21,363 & 91,286 & 8.55 & -3.35 & 0.13 & 0.0279 & 0.0479 & 0.4659 & 1.1788 & 0.8000 & NetRes \\
    ca-FluidDynamics & Collab. & 7,252 & 18,514 & 5.11 & -3.12 & 0.01 & 0.0801 & 0.1001 & 0.5367 & 1.3580 & 1.3400 & APS \\
    ca-GenTheoFiledParti & Collab. & 10,949 & 30,364 & 5.55 & -5.55 & 0.18 & 0.0428 & 0.0828 & 2.9702 & 7.5145 & 4.5000 & APS \\
    ca-GrQc & Collab. & 4,158 & 13,422 & 6.46 & -2.04 & 0.64 & 0.0225 & 0.0825 & 1.1689 & 2.9573 & 2.5200 & NetRes \\
    ca-HepPh & Collab. & 11,204 & 117,619 & 21.00 & -2.08 & 0.63 & 0.0041 & 0.0041 & 0.2463 & 0.6231 & 0.3200 & NetRes \\
    ca-InterdisPhysics & Collab. & 2,360 & 5,745 & 4.87 & -3.97 & -0.01 & 0.0840 & 0.1640 & 0.8638 & 2.1854 & 2.0000 & APS \\
    ca-MaterialsSci & Collab. & 12,438 & 41,912 & 6.74 & -3.98 & 0.08 & 0.0613 & 0.1013 & 2.3635 & 5.9797 & 4.6000 & APS \\
    ca-MatMetPhy & Collab. & 6,043 & 13,280 & 4.40 & -4.29 & 0.01 & 0.0782 & 0.1982 & 3.1866 & 8.0621 & 8.0000 & APS \\
    ca-NetSci2019 & Collab. & 32,904 & 296,876 & 18.05 & -2.02 & 0.99 & 0.0025 & 0.0425 & 0.5019 & 1.2699 & 0.7000 & ICN \\
    ca-NuclearAstroPhys & Collab. & 1,500 & 7,715 & 10.29 & -3.67 & 0.18 & 0.0354 & 0.0354 & 0.2746 & 0.6946 & 0.8500 & APS \\
    ca-StaPhyNonDyn & Collab. & 24,732 & 68,793 & 5.56 & -4.33 & 0.02 & 0.0433 & 0.0833 & 4.2011 & 10.6287 & 4.3000 & APS \\
    ca-Superconductivity & Collab. & 20,161 & 118,150 & 11.72 & -3.67 & 0.07 & 0.0210 & 0.0210 & 0.2199 & 0.5564 & 0.4700 & APS \\
    ca-TherProCondMat & Collab. & 2,328 & 8,613 & 7.40 & -3.36 & 0.08 & 0.0609 & 0.0809 & 0.3786 & 0.9578 & 1.4300 & APS \\
    cit-citeseer-2014 & Cite & 365,154 & 1,721,981 & 9.43 & -2.72 & -0.06 & 0.0190 & 0.0390 & 0.7111 & 1.7992 & 1.9000 & ICN \\
    cit-Database-System & Cite & 1,350 & 4,055 & 6.01 & -3.96 & 0.21 & 0.0787 & 0.0587 & 0.3131 & 0.7921 & 0.8200 & Aminer \\
    cit-DBLP & Cite & 12,495 & 49,563 & 7.93 & -3.35 & -0.05 & 0.0263 & 0.0263 & 0.2745 & 0.6946 & 0.5800 & NetRes \\
    cit-HepPh & Cite & 28,045 & 3,148,414 & 224.53 & -4.49 & 0.63 & 0.0014 & 0.0014 & 0.0080 & 0.0202 & 0.0140 & NetRes \\
    cit-HepTh & Cite & 22,721 & 2,444,642 & 215.19 & -3.66 & -0.03 & 0.0011 & 0.0011 & 0.0086 & 0.0218 & 0.0140 & NetRes \\
    email-dnc & Commun. & 1,833 & 4,366 & 4.76 & -1.86 & -0.31 & 0.0297 & 0.0297 & 0.4747 & 1.2010 & 1.2400 & NetRes \\
    email-enron-large & Commun. & 33,696 & 180,811 & 10.73 & -1.97 & -0.12 & 0.0087 & 0.0087 & 0.3048 & 0.7712 & 0.4500 & NetRes \\
    email-EU & Commun. & 32,430 & 54,397 & 3.35 & -4.25 & -0.38 & 0.0190 & 0.0190 & 0.9989 & 2.5273 & 1.4900 & NetRes \\
    email-EuAll & Commun. & 224,832 & 339,925 & 3.02 & -2.78 & -0.19 & 0.0103 & 0.0103 & 1.3831 & 3.4992 & 1.5600 & SNAP \\
    emails & Commun. & 56,576 & 92,013 & 3.25 & -1.89 & -0.08 & 0.0155 & 0.0155 & 0.8091 & 2.0471 & 2.6300 & Barabasi \\
    phonecalls & Commun. & 30,420 & 52,841 & 3.47 & -4.71 & 0.17 & 0.0768 & 0.1768 & 4.5878 & 11.6071 & 20.0000 & Barabasi \\
    cs4 & Dynamic & 22,499 & 43,858 & 3.90 & -3.27 & 0.32 & 0.3358 & 0.3758 & 0.2706 & 0.6845 & 1.8000 & NetRes \\
    ia-digg-reply & Dynamic & 29,652 & 84,781 & 5.72 & -3.52 & 0.00 & 0.0362 & 0.0362 & 0.3756 & 0.9502 & 0.7700 & NetRes \\
    ia-reality & Dynamic & 6,809 & 7,680 & 2.26 & -3.24 & -0.68 & 0.0858 & 0.0858 & 0.5818 & 1.4719 & 2.4000 & NetRes \\
    econ-poli & Econimic & 2,343 & 2,667 & 2.28 & -2.17 & -0.34 & 0.2310 & 0.3910 & 1.4633 & 3.7021 & 2.4000 & NetRes \\
    inf-power & Infras. & 4,941 & 6,594 & 2.67 & -3.22 & 0.00 & 0.1606 & 0.3606 & 3.7297 & 9.4361 & 17.0000 & NetRes \\
    power-bcspwr10 & Infras. & 5,300 & 8,271 & 3.12 & -8.58 & -0.05 & 0.2160 & 0.4160 & 5.0446 & 12.7629 & 13.4000 & NetRes \\
    road-minnesota & Infras. & 2,640 & 3,302 & 2.50 & -4.80 & -0.19 & 0.4920 & 0.6320 & 1.7043 & 4.3118 & 2.5000 & NetRes \\ 
    deezer-europe-edges & Social & 28,281 & 92,752 & 6.56 & -4.86 & 0.10 & 0.0435 & 0.0635 & 1.0388 & 2.6282 & 1.8000 & Github \\
    facebook-combined & Social & 4,039 & 88,234 & 43.69 & -2.51 & 0.06 & 0.0062 & 0.0062 & 0.0904 & 0.2288 & 0.1700 & SNAP \\
    lastfm-asia-edges & Social & 7,624 & 27,806 & 7.29 & -3.24 & 0.02 & 0.0272 & 0.0472 & 0.7767 & 1.9650 & 1.0300 & Github \\
    musae-DE-edges & Social & 9,498 & 153,138 & 32.25 & -2.55 & -0.12 & 0.0069 & 0.0069 & 0.0482 & 0.1220 & 0.1300 & SNAP \\
    musae-facebook & Social & 22,470 & 170,823 & 15.20 & -3.19 & 0.08 & 0.0097 & 0.0097 & 0.3259 & 0.8246 & 0.6000 & SNAP \\
    soc-fb-Huawei & Social & 1,000 & 50,153 & 100.31 & -7.69 & -0.01 & 0.0100 & 0.0100 & 0.0009 & 0.0023 & 0.0260 & Kaggle \\
    soc-musae-git & Social & 37,700 & 289,003 & 15.33 & -2.54 & -0.08 & 0.0074 & 0.0074 & 0.1218 & 0.3083 & 0.3400 & SNAP \\
    soc-instagram-Huawei & Social & 1,000 & 4,933 & 9.87 & -12.02 & 0.01 & 0.1015 & 0.0615 & 0.0337 & 0.0853 & 0.2300 & Kaggle \\
    soc-academia & Social & 200,167 & 1,022,440 & 10.22 & -3.11 & -0.02 & 0.0091 & 0.0091 & 0.3387 & 0.8569 & 1.3000 & NetRes \\
    soc-advogato & Social & 5,054 & 39,374 & 15.58 & -3.35 & -0.10 & 0.0148 & 0.0148 & 0.1132 & 0.2864 & 0.2800 & NetRes \\
    soc-delicious & Social & 536,108 & 1,365,961 & 5.10 & -2.80 & -0.07 & 0.0122 & 0.0322 & 1.0308 & 2.6079 & 1.6200 & NetRes \\
    soc-digg & Social & 770,799 & 5,907,132 & 15.33 & -1.70 & -0.09 & 0.0015 & 0.0015 & 0.4968 & 1.2568 & 0.5300 & NetRes \\
    soc-fb-pages-artist & Social & 50,515 & 819,090 & 32.43 & -3.06 & -0.02 & 0.0057 & 0.0057 & 0.0844 & 0.2136 & 0.1500 & NetRes \\
    soc-fb-pages-company & Social & 14,113 & 52,126 & 7.39 & -3.41 & 0.01 & 0.0327 & 0.0527 & 1.7180 & 4.3466 & 0.8700 & NetRes \\
    soc-fb-pages-government & Social & 7,057 & 89,429 & 25.34 & -3.06 & 0.03 & 0.0097 & 0.0097 & 0.1065 & 0.2693 & 0.2300 & NetRes \\
    soc-fb-pages-media & Social & 27,917 & 205,964 & 14.76 & -3.38 & 0.02 & 0.0168 & 0.0168 & 0.1528 & 0.3867 & 0.3600 & NetRes \\
    soc-Slashdot081106 & Social & 77,360 & 469,180 & 12.13 & -3.51 & -0.07 & 0.0078 & 0.0078 & 0.2810 & 0.7110 & 0.3300 & NetRes \\
    soc-twitter-Huawei & Social & 1,000 & 250,315 & 500.63 & -9.45 & 0.00 & 0.0020 & 0.0020 & 0.0001 & 0.0002 & 0.0046 & Kaggle \\
    tech-as-caida2007 & Techno. & 26,475 & 53,381 & 4.03 & -2.09 & -0.19 & 0.0168 & 0.0168 & 0.5819 & 1.4721 & 1.3200 & NetRes \\
    tech-p2p-gnutella & Techno. & 62,561 & 147,878 & 4.73 & -4.80 & -0.09 & 0.0871 & 0.0871 & 0.3918 & 0.9912 & 0.6100 & NetRes \\
    tech-pgp & Techno. & 10,680 & 24,316 & 4.55 & -4.26 & 0.24 & 0.0244 & 0.0444 & 1.7170 & 4.3440 & 2.6000 & NetRes \\
    web-EPA & Web & 4,253 & 8,897 & 4.18 & -2.49 & -0.30 & 0.0725 & 0.0725 & 0.3807 & 0.9631 & 0.9800 & NetRes \\
    web-webbase-2001 & Web & 16,062 & 25,593 & 3.19 & -2.09 & -0.10 & 0.0323 & 0.0523 & 6.8700 & 17.3811 & 14.0000 & NetRes
\end{longtable}
}

\subsection{Network models}

We also generate a series of scale-free networks using different models.
First, we create degree sequences by setting the degree exponent~($\gamma$) and minimum degree~($k_{min}$).
Then we feed the sequence to the configuration model to create the networks while forbidding self-loops and multi-links.
Since the edge density of the networks generated this way is very sensitive to the choice of $\gamma$, we set it to 3.3 for all instances and tune the edge density using $k_{min}$.
To generate scale-free networks with varying $\gamma$ values, we use the Goh-Kahng-Kim~(GKK) algorithm~\cite{Goh2001}.
The edge density of the networks generated by it is less sensitive to $\gamma$.
We also apply the Xalvi-Brunet $\&$ Sokolov algorithm~\cite{Xulvi2004} to an instance of the networks generated by the configuration model to produce a series of scale-free networks with different assortativeness.
The model has a key parameter $p$ that defines the edge rewiring probability.

The parameters and the statistics of all synthetic networks can be found in Supplementary Table.~\ref{tab:Tab2}.
Since these networks are mainly used to fit the relation between $\alpha_c$ and $\mathcal{L}$, we only report the predicted values of $\beta_c$ and numeric values of $\alpha_c$.

\setlength{\tabcolsep}{5.0pt}
\setlength{\LTcapwidth}{\textwidth}
{
\small
\begin{longtable}{lrrrrrrrr}
    \caption{
    Properties of the network models.
    We report the generating method of the network, key parameter, number of nodes $N$, number of edges $L$, average degree $\langle k \rangle$, degree exponent $\gamma$, degree correlation coefficient $r$, predicted critical point $\beta_c$, localization strength $\mathcal{L}$, numerical tricritical point $\alpha_c^{num}$.} 
    \label{tab:Tab2}
    \\\hline
    \multicolumn{1}{c}{Network} &  \multicolumn{1}{c}{$N$} &  \multicolumn{1}{c}{$L$} &  \multicolumn{1}{c}{$\langle k \rangle$} &  \multicolumn{1}{c}{$\gamma$} &  \multicolumn{1}{c}{$r$} &  \multicolumn{1}{c}{$\beta_c$} &  \multicolumn{1}{c}{$\mathcal{L}$} &  \multicolumn{1}{c}{$\alpha_c^{num}$} \\
    \hline
    \endfirsthead
     
    \\\hline 
    \multicolumn{1}{c}{Network} &  \multicolumn{1}{c}{$N$} &  \multicolumn{1}{c}{$L$} &  \multicolumn{1}{c}{$\langle k \rangle$} &  \multicolumn{1}{c}{$\gamma$} &  \multicolumn{1}{c}{$r$} &  \multicolumn{1}{c}{$\beta_c$} &  \multicolumn{1}{c}{$\mathcal{L}$} &  \multicolumn{1}{c}{$\alpha_c^{num}$} 
    \\\hline
    \endhead     

    \hline 
    \multicolumn{3}{l}{{Continued on next page}} \\ 
    \endfoot

    \hline
    \endlastfoot
    CM ($k_{min}=1$) & 100,000 & 137,836 & 2.76 & 3.3 & 0.0006 & 0.2677 & 0.6572 & 1.48 \\
    CM ($k_{min}=2$) & 100,000 & 225,218 & 4.50 & 3.3 & -0.0013 & 0.1353 & 0.3335 & 0.92 \\
    CM ($k_{min}=3$) & 100,000 & 313,560 & 6.27 & 3.3 & -0.0033 & 0.0876 & 0.2255 & 0.72 \\
    CM ($k_{min}=4$) & 100,000 & 399,294 & 7.99 & 3.3 & -0.0017 & 0.0717 & 0.1468 & 0.45 \\
    CM ($k_{min}=5$) & 100,000 & 492,157 & 9.84 & 3.3 & -0.0030 & 0.0504 & 0.1325 & 0.46 \\
    CM ($k_{min}=6$)  & 100,000 & 577,430 & 11.55 & 3.3 & -0.0025 & 0.0481 & 0.0953 & 0.31 \\
    CM ($k_{min}=7$) & 100,000 & 660,324 & 13.21 & 3.3 & 0.0032 & 0.0438 & 0.0781 & 0.25 \\
    CM ($k_{min}=8$) & 100,000 & 752,652 & 15.05 & 3.3 & -0.0016 & 0.0357 & 0.0728 & 0.26 \\
    CM ($k_{min}=9$) & 100,000 & 837,560 & 16.75 & 3.3 & -0.0020 & 0.0334 & 0.0615 & 0.20 \\
    CM (Disassort, $p=1.0$) & 100,000 & 138,849 & 2.78 & 3.3 & -0.0540 & 0.3248 & 0.6789 & 1.30 \\
    CM (Disassort, $p=0.9$) & 100,000 & 138,849 & 2.78 & 3.3 & -0.0484 & 0.3214 & 0.6997 & 1.32 \\
    CM (Disassort, $p=0.8$) & 100,000 & 138,849 & 2.78 & 3.3 & -0.0436 & 0.3193 & 0.6682 & 1.22 \\
    CM (Disassort, $p=0.7$) & 100,000 & 138,849 & 2.78 & 3.3 & -0.0400 & 0.3078 & 0.6870 & 1.39 \\
    CM (Disassort, $p=0.6$) & 100,000 & 138,849 & 2.78 & 3.3 & -0.0345 & 0.3010 & 0.6804 & 1.36 \\
    CM (Disassort, $p=0.5$) & 100,000 & 138,849 & 2.78 & 3.3 & -0.0316 & 0.2945 & 0.6890 & 1.41 \\
    CM (Disassort, $p=0.4$) & 100,000 & 138,849 & 2.78 & 3.3 & -0.0213 & 0.2789 & 0.7194 & 1.62 \\
    CM (Disassort, $p=0.3$) & 100,000 & 138,849 & 2.78 & 3.3 & -0.0190 & 0.2714 & 0.6962 & 1.60 \\
    CM (Disassort, $p=0.2$) & 100,000 & 138,849 & 2.78 & 3.3 & -0.0158 & 0.2688 & 0.6436 & 1.50 \\
    CM (Disassort, $p=0.1$) & 100,000 & 138,849 & 2.78 & 3.3 & -0.0043 & 0.2300 & 0.8707 & 2.48 \\
    CM (Assort, $p=0.1$) & 100,000 & 138,849 & 2.78 & 3.3 & 0.0088 & 0.2087 & 0.8262 & 2.44 \\
    CM (Assort, $p=0.2$)& 100,000 & 138,849 & 2.78 & 3.3 & 0.0160 & 0.1946 & 0.8376 & 2.50 \\
    CM (Assort, $p=0.3$) & 100,000 & 138,849 & 2.78 & 3.3 & 0.0351 & 0.1634 & 1.0429 & 3.46 \\
    CM (Assort, $p=0.4$) & 100,000 & 138,849 & 2.78 & 3.3 & 0.0435 & 0.1519 & 1.0179 & 3.66 \\
    CM (Assort, $p=0.5$) & 100,000 & 138,849 & 2.78 & 3.3 & 0.0596 & 0.1333 & 1.1410 & 4.34 \\
    CM (Assort, $p=0.6$) & 100,000 & 138,849 & 2.78 & 3.3 & 0.0769 & 0.1165 & 1.2780 & 5.04 \\
    CM (Assort, $p=0.7$) & 100,000 & 138,849 & 2.78 & 3.3 & 0.0860 & 0.1070 & 1.3438 & 5.20 \\
    CM (Assort, $p=0.8$) & 100,000 & 138,849 & 2.78 & 3.3 & 0.0966 & 0.0988 & 1.4131 & 5.35 \\
    CM (Assort, $p=0.9$) & 100,000 & 138,849 & 2.78 & 3.3 & 0.1170 & 0.0884 & 1.5549 & 5.70 \\
    CM (Assort, $p=1.0$) & 100,000 & 138,849 & 2.78 & 3.3 & 0.1410 & 0.0770 & 1.7690 & 6.50 \\
    GKK ($\gamma=2.1$) & 51,490 & 100,000 & 3.88 & 2.1 & -0.1108 & 0.0109 & 0.7665 & 1.50 \\
    GKK ($\gamma=2.2$) & 57,539 & 100,000 & 3.48 & 2.2 & -0.0783 & 0.0147 & 0.8748 & 2.25 \\
    GKK ($\gamma=2.3$) & 62,597 & 100,000 & 3.20 & 2.3 & -0.0542 & 0.0201 & 0.9692 & 3.25 \\
    GKK ($\gamma=2.4$) & 66,554 & 100,000 & 3.01 & 2.4 & -0.0345 & 0.0282 & 1.0284 & 3.26 \\
    GKK ($\gamma=2.5$) & 69,853 & 100,000 & 2.86 & 2.5 & -0.0208 & 0.0415 & 1.0476 & 3.56 \\
    GKK ($\gamma=2.6$) & 72,258 & 100,000 & 2.77 & 2.6 & -0.0123 & 0.0572 & 1.0701 & 3.81 \\
    GKK ($\gamma=2.7$) & 74,223 & 100,000 & 2.69 & 2.7 & -0.0111 & 0.0849 & 1.0025 & 3.41 \\
    GKK ($\gamma=2.8$) & 75,961 & 100,000 & 2.63 & 2.8 & -0.0030 & 0.1068 & 1.0280 & 3.28 \\
    GKK ($\gamma=2.9$) & 77,173 & 100,000 & 2.59 & 2.9 & -0.0051 & 0.1403 & 0.9539 & 2.8 \\
    GKK ($\gamma=3.0$) & 78,203 & 100,000 & 2.56 & 3.0 & 0.0022 & 0.1635 & 0.9798 & 2.68 \\
    GKK ($\gamma=3.1$) & 79,005 & 100,000 & 2.53 & 3.1 & -0.0010 & 0.2005 & 0.8992 & 2.16 \\
    GKK ($\gamma=3.2$) & 79,656 & 100,000 & 2.51 & 3.2 & 0.0003 & 0.2299 & 0.8220 & 1.74 \\
    GKK ($\gamma=3.3$) & 80,304 & 100,000 & 2.49 & 3.3 & -0.0074 & 0.2685 & 0.6786 & 1.32 \\
    GKK ($\gamma=3.4$) & 80,884 & 100,000 & 2.47 & 3.4 & 0.0026 & 0.2774 & 0.7720 & 1.34 \\
    GKK ($\gamma=3.5$) & 81,390 & 100,000 & 2.46 & 3.5 & -0.0005 & 0.3082 & 0.6222 & 1.17 \\
    GKK ($\gamma=3.6$) & 81,789 & 100,000 & 2.45 & 3.6 & 0.0033 & 0.3223 & 0.6329 & 1.11 \\
    GKK ($\gamma=3.7$) & 82,045 & 100,000 & 2.44 & 3.7 & -0.0002 & 0.3392 & 0.6133 & 1.05 \\
    GKK ($\gamma=3.8$) & 82,637 & 100,000 & 2.42 & 3.8 & 0.0007 & 0.3560 & 0.5437 & 1.00 \\
    GKK ($\gamma=3.9$) & 82,819 & 100,000 & 2.41 & 3.9 & -0.0023 & 0.3718 & 0.5233 & 0.92 \\
    GKK ($\gamma=4.0$) & 83,041 & 100,000 & 2.41 & 4.0 & -0.0017 & 0.3840 & 0.4924 & 0.89 \\
    \hline
\end{longtable}
}

\section{Supplementary results}

\subsection{Implications of the tricritical point}

\begin{figure}[!htbp]
    \centering
    \includegraphics[width=1.0\columnwidth]{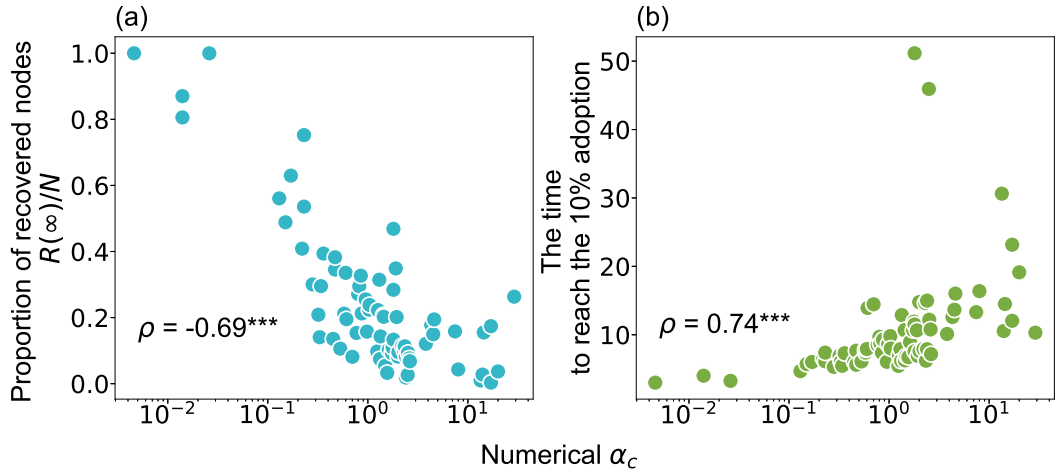}
    \caption{
    The implications of tricritical point $\alpha_c$ for real-world networks.
    (a) The final proportion of recovered nodes versus $\alpha_c$.
    (b) The time needed for $10\%$ individuals to adopt the product versus $\alpha_c$.
    Each marker corresponds to an empirical network and the result is the average of 10,000 simulations with the parameter ($\beta_c$+0.1, $\alpha_c$).
    The Spearman's correlation coefficients $\rho$ are annotated in the figure.
    Both cases are statistically significant at the level 0.001.
    }
    \label{Sfig:meaning_realnet}
\end{figure}

By definition, $\alpha$ is the intensity of the macro-level influence and $\alpha_c$ determines the phase transition class of the diffusion process.
Here we show that $\alpha_c$ has further implications.
Through simulations on the real-world networks, we find that networks with larger $\alpha_c$ values tend to have lower market penetration levels (Spearman's $\rho=-0.69, p<0.001$, see Supplementary Figure~\ref{Sfig:meaning_realnet}(a)).
We also examine the time required for $10\%$ of the individuals to adopt the product and find a positive association between it and $\alpha_c$ (Spearman's $\rho=0.74, p<0.001$, see Supplementary Figure~\ref{Sfig:meaning_realnet}(b)).
Given these correlations, a smaller $\alpha_c$ value is desirable for marketers.

\subsection{The effect of macro-level influence on $\beta_c$}

\begin{figure}[!htbp]
    \centering
    \includegraphics[width=1.0\columnwidth]{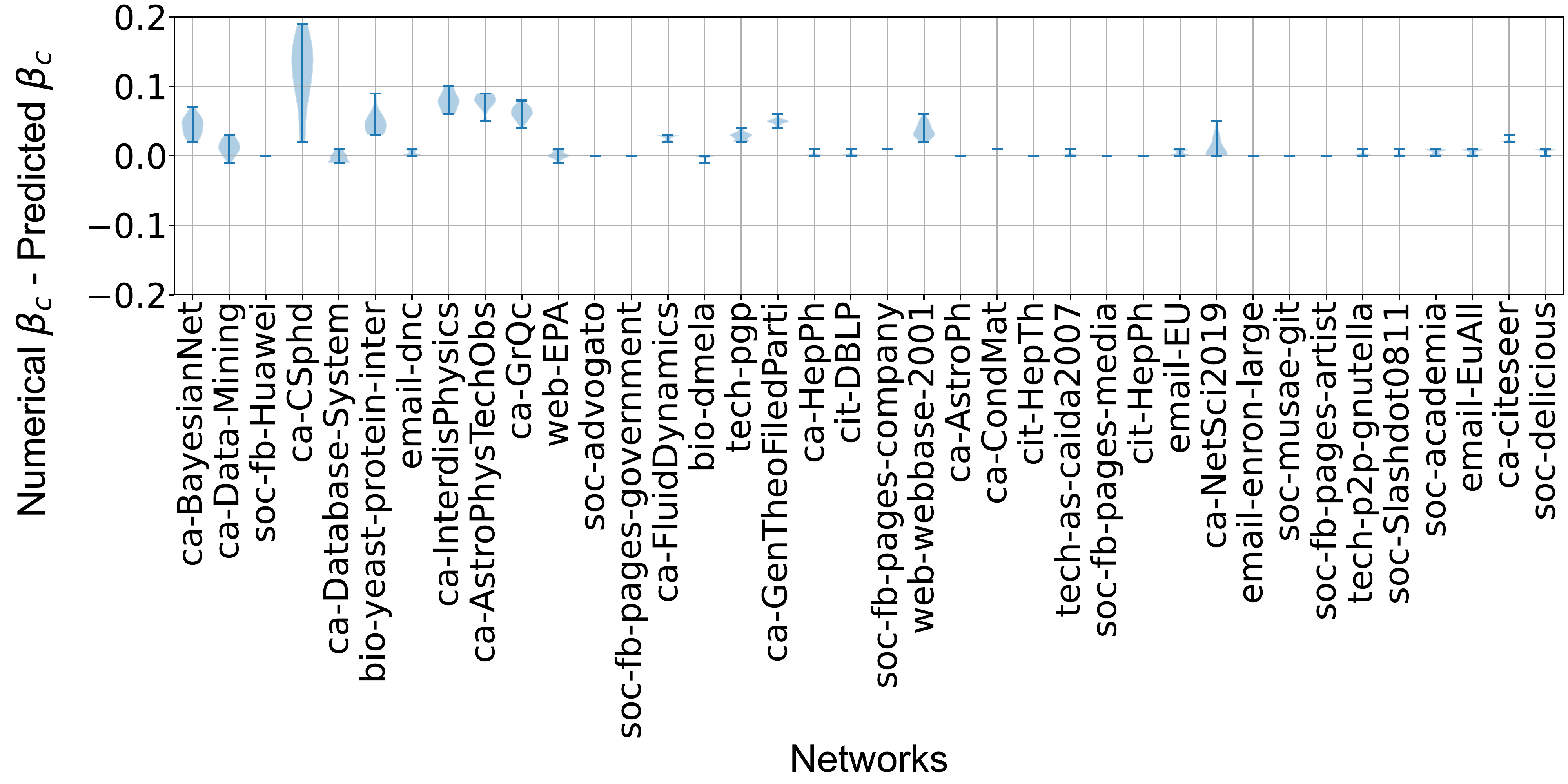}
    \caption{
    Discrepancy between the numerical and theoretical values of $\beta_c$.
    For each network, we run the simulations using different $\alpha$ values ranging from 0 to 1 and identify the $\beta_c$ values.
    We then plot the distribution of the discrepancy between the numerical values and the theoretical value.
    The size of the network increases monotonically from left to right.}
    \label{Sfig:criticalPoint}
\end{figure}

In the main text, we state that the introduction of the macro-level influence does not affect the value of $\beta_c$.
We perform numerous simulations with varying $\alpha$ values on 38 empirical networks randomly selected from our collection to confirm this.
We show the discrepancy between the numerical and theoretical values of $\beta_c$ in Supplementary Figure~\ref{Sfig:criticalPoint}.
It appears that macro-level influence leads to $\beta_c$ values that are larger than expected in some cases, but the difference decreases as the size of the network increases.
This suggests that the finite-size effect is at play.

\subsection{Other approaches to quantify the localization strength}
\label{Comparision: localization metrics}

\begin{figure}[!htbp]
    \centering
    \includegraphics[width=1.0\columnwidth]{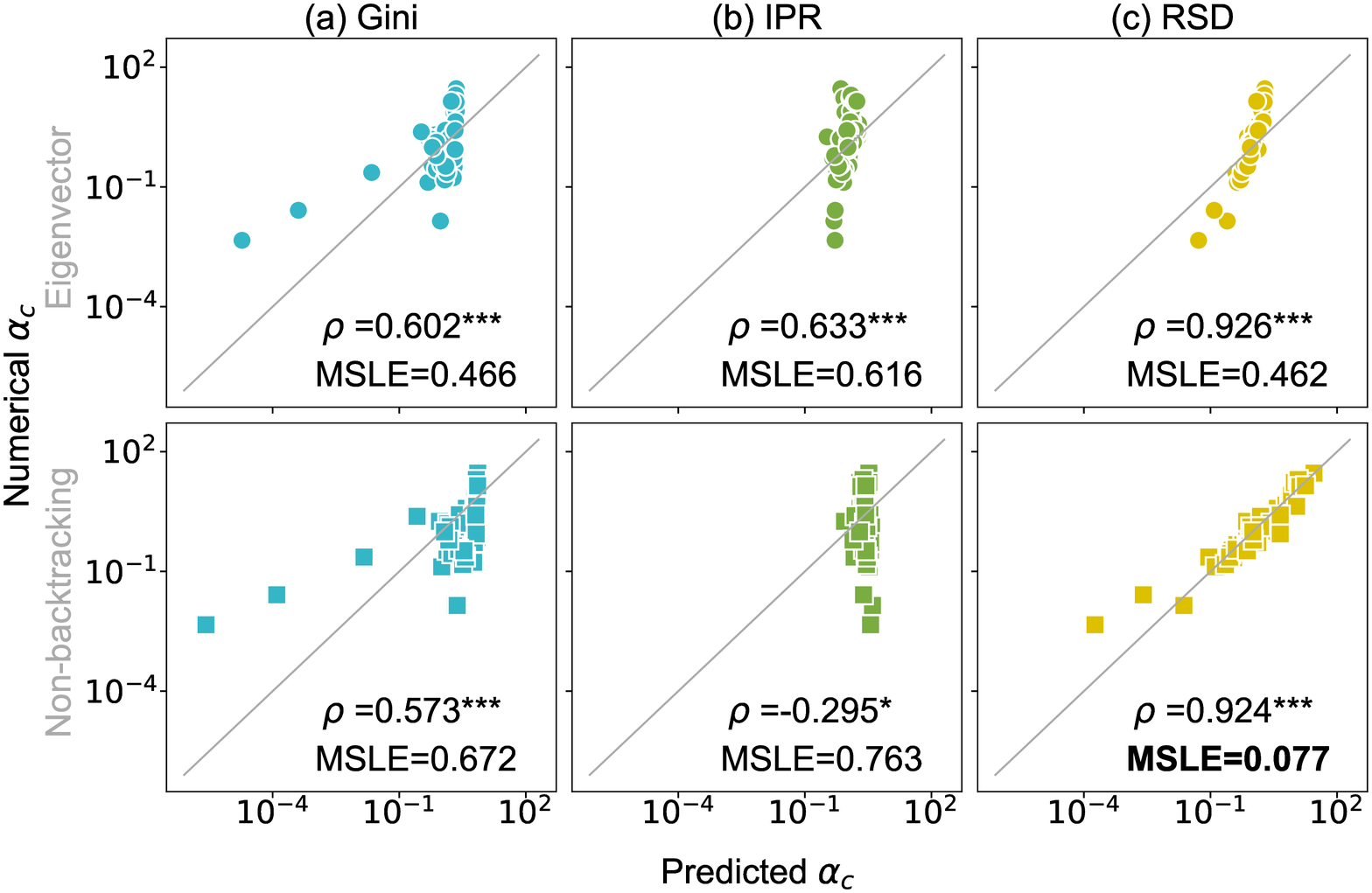}
    \caption{
    Comparison of different methods to predict the tricritical point.
    We combine the eigenvector and non-backtracking centrality measures with (a) Gini coefficient, (b) IPR, and (c) re-scaled standard deviation (RSD) to quantify the localization phenomenon of networks, respectively. 
    Spearman's correlation coefficient $\rho$ and the mean squared logarithmic errors (MSLE)  between the numerical and predicted values of $\alpha_c$  are annotated in the plots~(***p<0.001,**p<0.01,*p<0.05). 
    The gray solid line $y = x$ is added to guide the eye. 
    }
    \label{Sfig:metrics}
\end{figure}

Our main contribution is to show that the tricritical point is associated with the localization effect.
Network localization refers to the phenomenon that the components $x_i$ of an eigenvector are concentrated on a subset of nodes. 
Considerable efforts have been devoted to studying the localization of the eigenvector and non-backtracking centrality due to their fundamental role in spreading dynamics.
A common metric in the literature is inverse participation ratio (IPR), defined as $IPR=\sum_{i} x_i^4$.
Instead of re-scaled standard deviation (RSD) proposed in the present work, one can also use the Gini coefficient to quantify the variance of $x_i$.

To show that our method works better in predicting the values of $\alpha_{c}$, we compare the metric with other approaches.
Here we consider two approaches to calculate node centrality: eigenvector and non-backtracking centrality and three approaches to measure the variance: IPR, RSD, and Gini coefficient, which lead to six different metrics.
We fit their values against the numerical values of $\alpha_c$ on the network models to calculate the coefficients, then calculate the predicted values of $\alpha_c$ for the real-world networks.
Two methods are adopted to evaluate the performance of different localization measures.
First, we calculate Spearman's correlation coefficients between the numerical and predicted values.
Second, since the $\alpha_c$ values for different networks have different magnitudes, we calculate the mean squared logarithmic error (MSLE).
It is defined as
\begin{equation}
    E = \frac{1}{m} \sum^{m}_{j=1} (\log (\alpha_{c,j}^{num}+1) - \log (\alpha_{c,j}+1))^2,
\end{equation}
where $\alpha_{c,j}^{num}$ and $\alpha_{c,j}$ stand for the numerical and predicted tricritical points for network $j$.

We show the results in Supplementary Figure~\ref{Sfig:metrics}.
We can see that all predicted values are positively correlated with the numerical values except for IPR of the non-backtracking centrality. 
The standard deviation approaches, both leveraging eigenvector and non-backtracking centrality, yield very high correlation coefficients.
However, the one based on non-backtracking centrality has a much smaller MSLE.
This comparison confirms that our proposed method is more accurate than other approaches.

\subsection{Identifying the class of phase transition}

\begin{figure}[!htbp]
    \centering
    \includegraphics[width=1.0\columnwidth]{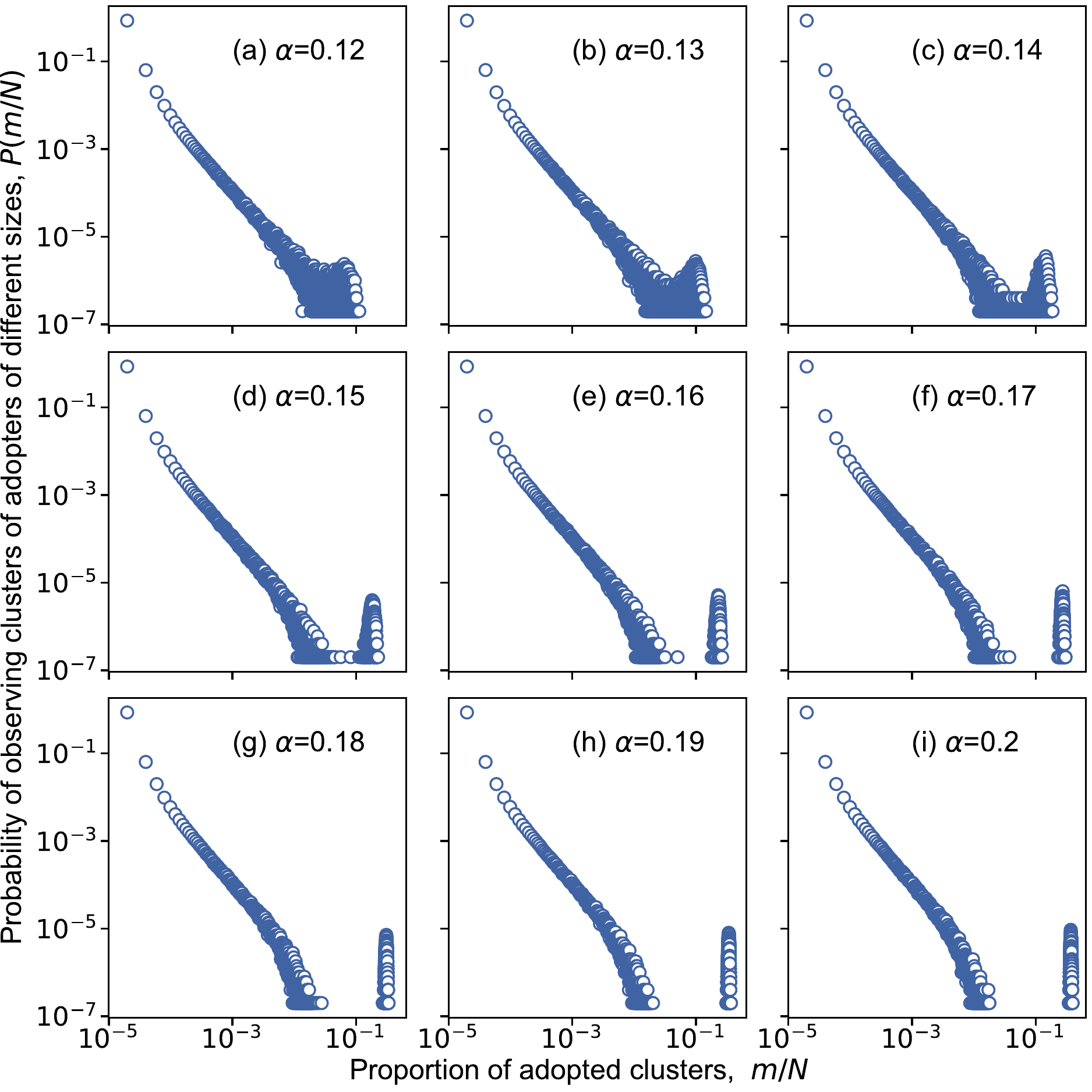}
    \caption{
    Probabilities of observing adopter clusters of different sizes with different $\alpha$ values on the soc-fb-pages-artist network. 
    For a pair of parameters~($\beta_c$,$\alpha$), we perform $5\times10^6$ simulations to calculate the probability of forming the clusters with different sizes.
    $m$ denotes the size of adopter clusters. 
    }
    \label{Sfig:massdistribution}
\end{figure}

To identify the numerical tricritical point, we plot the distribution of probabilities of observing adopter clusters of different sizes at $\beta_c$ while varying the value of $\alpha$.
See Supplementary Figure~\ref{Sfig:massdistribution} for the examples on the soc-fb-pages-artist network.
The dots that follow a power-law distribution in the figure correspond to the premature clusters of adopters that die out in the early stage of diffusion.
When $\alpha$ is close to 0.15, a spike of mass distribution on the right-hand side starts to emerge.
These dots represent the giant clusters of adopted individuals, which occupy a finite fraction of the system.
The phase transition is discontinuous when the giant clusters are well separated from the premature ones and this is how we determine the critical value $\alpha_c$

\subsection{Relation between predicted critical and tricritical point}

\begin{figure}[!htbp]
    \centering
    \includegraphics[width=1.0\columnwidth]{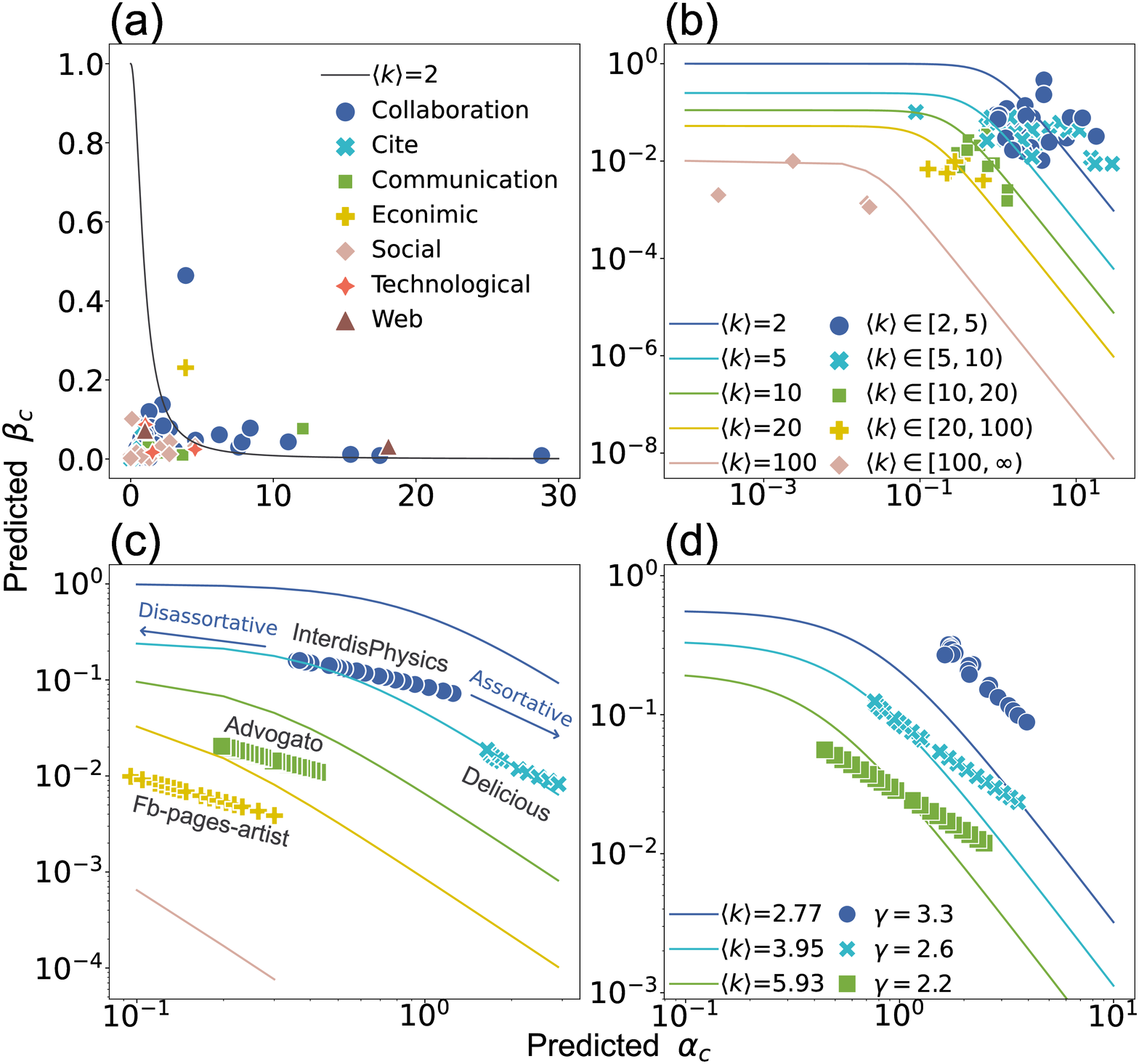}
    \caption{\textbf{Relationship between the critical and tricritical points}. 
    $\beta_{c}$ and $\alpha_{c}$ are obtained theoretically from Eq.~(\ref{eq:eqs23}) and Eq.~\ref{eq:alpha_l}, respectively.
    Each dot corresponds to a network and its location indicates the critical and tricritical values obtained from predicted results.
    The solid line represents the relation between $\beta_c$ and $\alpha_c$ described by Eq.~\ref{eq:beta_alpha_relation} with a  given mean degree.
    (a) We mark the categories of the networks and show the line with $\langle k \rangle = 2$.
    (b) We mark the networks according to their average degree and show the lines corresponding to different $\langle k \rangle$ values.
    The plot is in log-scale to highlight the details.
    (c) By changing the network's assortativeness, we obtain configurations with different $\beta_c$ and $\alpha_c$ values for four selected real-world networks~(i.e., InterdisPhysics, Delicious, Advogato, Fb-pages-artist).
    The directions of the changes are annotated in the plot.
    (d) Same as (c) but with three instances of scale-free networks generated by the configuration model.  
    }
    \label{Sfig:relation_numerical}
\end{figure}

To examine the relation between $\beta_c$ and $\alpha_c$, we show the scatter plot of their numerical values obtained from different real-world networks in the main text.
Here we show their predicted values in Supplementary figure~\ref{Sfig:relation_numerical}(a), the patterns are qualitatively similar to those in the main text.

\subsection{Tuning the localization strength of network}

\begin{figure}[!htbp]
    \centering
    \includegraphics[width=1.0\columnwidth]{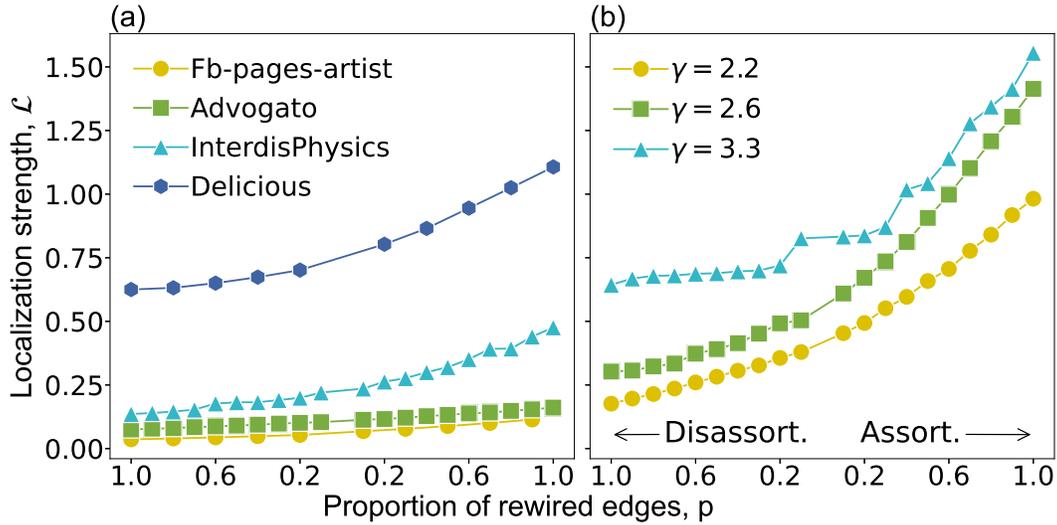}
    \caption{
    Varying the localization strength of networks by tuning the magnitude of degree correlation.
    We apply the Xalvi-Brunet $\&$ Sokolov algorithm with different portions of rewired edges on (a) four real-world networks~(Fb-pages-artist, Advogato, InterdisPhysics, Delicious) and (b) three instances of scale-free networks generated by configuration model with minimum degree $k=1$ and different $\gamma$ values.
    }
    \label{Sfig:localstength_assort}
\end{figure}

We present an approach to tune the localization strength of the network by changing the degree correlation through the Xalvi-Brunet $\&$ Sokolov algorithm.
It works by selecting two edges at random and ordering the four nodes at the ends of the edges according to their degree for rewiring.
If the goal is to increase the assortativeness, the algorithm connects the two nodes with the highest degree and the two with the lowest degree.
Conversely, the node with the highest degree is connected to the one with the lowest degree.
To tune the magnitude of degree correlation, we execute the deterministic rewiring step with probability $p$, and randomly pair the four nodes with the probability $1-p$.  
By fixing the total number of rewired steps, we can generate the network with different magnitudes of degree correlation by varying $p$.
The results for four real-world networks and three synthetic networks are shown in Supplementary Figure~\ref{Sfig:localstength_assort}.
The localization strength increases with the network assortativity level.


\begin{thebibliography}{10}

\bibitem{Rogers2003}
Everett~M Rogers.
\newblock {\em Diffusion of innovations}.
\newblock New York: The Free Press, 2003.

\bibitem{Chatterjee1990}
Rabik~Ar Chatterjee and Jehoshua Eliashberg.
\newblock The innovation diffusion process in a heterogeneous population: A
  micromodeling approach.
\newblock {\em Manage Sci}, 36(9):1057--1079, 1990.

\bibitem{Berry2018}
Frances~Stokes Berry and William~D Berry.
\newblock Innovation and diffusion models in policy research.
\newblock {\em Theories of the policy process}, pages 253--297, 2018.

\bibitem{Gabriel2014}
Gabriel~E. Kreindler and H.~Peyton Young.
\newblock Rapid innovation diffusion in social networks.
\newblock {\em Proc. Natl. Acad. Sci.}, 111:10881--10888, 2014.

\bibitem{Bass1969}
Frank~M Bass.
\newblock A new product growth for model consumer durables.
\newblock {\em Manage Sci}, 15(5):215--227, 1969.

\bibitem{Kiesling2012}
Elmar Kiesling, Markus G{\"u}nther, Christian Stummer, and Lea~M Wakolbinger.
\newblock Agent-based simulation of innovation diffusion: a review.
\newblock {\em Cent Eur J Oper Res}, 20(2):183--230, 2012.

\bibitem{Arts2011}
Joep~WC Arts, Ruud~T Frambach, and Tammo~HA Bijmolt.
\newblock Generalizations on consumer innovation adoption: A meta-analysis on
  drivers of intention and behavior.
\newblock {\em Int. J. Res. Mark.}, 28(2):134--144, 2011.

\bibitem{Jackson2010}
Matthew~O Jackson.
\newblock {\em Social and economic networks}.
\newblock Princeton university press, 2010.

\bibitem{Bale2013}
Catherine~SE Bale, Nicholas~J McCullen, Timothy~J Foxon, Alastair~M Rucklidge,
  and William~F Gale.
\newblock Harnessing social networks for promoting adoption of energy
  technologies in the domestic sector.
\newblock {\em Energy Policy}, 63:833--844, 2013.

\bibitem{Mccoy2014}
Daire McCoy and Se{\'a}n Lyons.
\newblock Consumer preferences and the influence of networks in electric
  vehicle diffusion: An agent-based microsimulation in ireland.
\newblock {\em Energy Res. Soc. Sci.}, 3:89--101, 2014.

\bibitem{Rahmandad2008}
Hazhir Rahmandad and John Sterman.
\newblock Heterogeneity and network structure in the dynamics of diffusion:
  Comparing agent-based and differential equation models.
\newblock {\em Manage Sci}, 54(5):998--1014, 2008.

\bibitem{Peres2010}
Renana Peres, Eitan Muller, and Vijay Mahajan.
\newblock Innovation diffusion and new product growth models: A critical review
  and research directions.
\newblock {\em Int. J. Res. Mark.}, 27(2):91--106, 2010.
\numberwithin{equation}{section}
\bibitem{Goldenberg2009}
Jacob Goldenberg, Sangman Han, Donald~R Lehmann, and Jae~Weon Hong.
\newblock The role of hubs in the adoption process.
\newblock {\em J Mark}, 73(2):1--13, 2009.

\bibitem{Delre2010}
Sebastiano~A Delre, Wander Jager, Tammo~HA Bijmolt, and Marco~A Janssen.
\newblock Will it spread or not? the effects of social influences and network
  topology on innovation diffusion.
\newblock {\em J Prod Innov Manage}, 27(2):267--282, 2010.

\bibitem{Castellano2009}
Claudio Castellano, Santo Fortunato, and Vittorio Loreto.
\newblock Statistical physics of social dynamics.
\newblock {\em Rev. Mod. Phys.}, 81:591--646, May 2009.

\bibitem{Pastor2015}
Romualdo Pastor-Satorras, Claudio Castellano, Piet Van~Mieghem, and Alessandro
  Vespignani.
\newblock Epidemic processes in complex networks.
\newblock {\em Rev. Mod. Phys.}, 87:925--979, Aug 2015.

\bibitem{Hohnisch2008}
Martin Hohnisch, Sabine Pittnauer, and Dietrich Stauffer.
\newblock A percolation-based model explaining delayed takeoff in new-product
  diffusion.
\newblock {\em Ind. Corp. Chang.}, 17(5):1001--1017, 2008.

\bibitem{Buttle1998}
Francis~A Buttle.
\newblock Word of mouth: understanding and managing referral marketing.
\newblock {\em J. Strateg. Mark.}, 6(3):241--254, 1998.

\bibitem{Young2009}
H~Peyton Young.
\newblock Innovation diffusion in heterogeneous populations: Contagion, social
  influence, and social learning.
\newblock {\em Am Econ Rev}, 99(5):1899--1924, 2009.

\bibitem{Chen2016}
Ching-Wen Chen, Hung-Yi Chang, Juin-Han Chen, and Richard Weng.
\newblock Elucidating the role of conformity in innovative smartphones.
\newblock {\em Int. J. Mob. Commun.}, 14(1):56--78, 2016.

\bibitem{Guilbeault2018}
Douglas Guilbeault, Joshu Becker, and Damon Centola.
\newblock {\em Complex Contagions: A Decade in Review}, pages 3--25.
\newblock Springer International Publishing, Cham, 2018.

\bibitem{Katz1992}
Michael~L. Katz and Carl Shapiro.
\newblock Product introduction with network externalities.
\newblock {\em J Ind Econ}, 40(1):55--83, 1992.

\bibitem{Schoder2000}
Detlef Schoder.
\newblock Forecasting the success of telecommunication services in the presence
  of network effects.
\newblock {\em Inf. Econ. Policy}, 12(2):181--200, 2000.

\bibitem{Rohlfs2003}
Jeffrey~H Rohlfs.
\newblock {\em Bandwagon effects in high-technology industries}.
\newblock MIT press, 2003.

\bibitem{Scherer1990}
Frederic~M Scherer and David Ross.
\newblock Industrial market structure and economic performance.
\newblock {\em University of Illinois at Urbana-Champaign's Academy for
  entrepreneurial leadership historical research reference in
  entrepreneurship}, 1990.

\bibitem{Browning2020}
Edgar~K Browning and Mark~A Zupan.
\newblock {\em Microeconomics: Theory and applications}.
\newblock John Wiley \& Sons, 2020.

\bibitem{Goltsev2012}
A.~V. Goltsev, S.~N. Dorogovtsev, J.~G. Oliveira, and J.~F.~F. Mendes.
\newblock Localization and spreading of diseases in complex networks.
\newblock {\em Phys. Rev. Lett.}, 109:128702, Sep 2012.

\bibitem{Travis2014}
Travis Martin, Xiao Zhang, and M.~E.~J. Newman.
\newblock Localization and centrality in networks.
\newblock {\em Phys. Rev. E}, 90:052808, Nov 2014.

\bibitem{Lechman2018}
Ewa Lechman.
\newblock Networks externalities as social phenomenon in the process ict
  diffusion.
\newblock {\em Econ. Sociol.}, 11:22--40, 2018.

\bibitem{Boccaletti2016}
S~Boccaletti, JA~Almendral, S~Guan, I~Leyva, Z~Liu, I~Sendi{\~n}a-Nadal,
  Z~Wang, and Y~Zou.
\newblock Explosive transitions in complex networks’ structure and dynamics:
  Percolation and synchronization.
\newblock {\em Phys. Rep.}, 660:1--94, 2016.

\bibitem{D2019}
Raissa~M D'Souza, Jesus G{\'o}mez-Gardenes, Jan Nagler, and Alex Arenas.
\newblock Explosive phenomena in complex networks.
\newblock {\em Adv. Phys.}, 68(3):123--223, 2019.

\bibitem{Koher2019}
Andreas Koher, Hartmut H.~K. Lentz, James~P. Gleeson, and Philipp H\"ovel.
\newblock Contact-based model for epidemic spreading on temporal networks.
\newblock {\em Phys. Rev. X}, 9:031017, Aug 2019.

\bibitem{Pastor2020}
Romualdo Pastor-Satorras and Claudio Castellano.
\newblock The localization of non-backtracking centrality in networks and its
  physical consequences.
\newblock {\em Sci. Rep.}, 10(1):1--12, 2020.

\bibitem{Cai2015}
Weiran Cai, Li~Chen, Fakhteh Ghanbarnejad, and Peter Grassberger.
\newblock {Avalanche outbreaks emerging in cooperative contagions}.
\newblock {\em Nat. Phys.}, 11(11):936--940, 2015.

\bibitem{Cui2018}
Peng-Bi Cui, Wei Wang, Shi-Min Cai, Tao Zhou, Ying-Cheng Lai, et~al.
\newblock Close and ordinary social contacts: How important are they in
  promoting large-scale contagion?
\newblock {\em Phys. Rev. E}, 98(5):052311, 2018.

\bibitem{Shrestha2015}
Munik Shrestha, Samuel~V Scarpino, and Cristopher Moore.
\newblock Message-passing approach for recurrent-state epidemic models on
  networks.
\newblock {\em Phys. Rev. E}, 92(2):022821, 2015.

\bibitem{Timar2022}
G~Tim{\'a}r, SN~Dorogovtsev, and JFF Mendes.
\newblock Localization of nonbacktracking centrality on dense subgraphs of
  sparse networks.
\newblock {\em arXiv preprint arXiv:2209.02594}, 2022.

\bibitem{Bonacich1972}
Phillip Bonacich.
\newblock Factoring and weighting approaches to status scores and clique
  identification.
\newblock {\em J Math Sociol}, 2(1):113--120, 1972.

\bibitem{Bendel1989}
RB~Bendel, SS~Higgins, JE~Teberg, and DA~Pyke.
\newblock Comparison of skewness coefficient, coefficient of variation, and
  gini coefficient as inequality measures within populations.
\newblock {\em Oecologia}, 78(3):394--400, 1989.

\bibitem{Sheng2013}
Margaret~L Sheng, Shen-Yao Chang, Thompson Teo, and Yuh-Feng Lin.
\newblock Knowledge barriers, knowledge transfer, and innovation competitive
  advantage in healthcare settings.
\newblock {\em Manag. Decis.}, 51(3):461--478, 2013.

\bibitem{Xulvi2004}
R.~Xulvi-Brunet and I.~M. Sokolov.
\newblock Reshuffling scale-free networks: From random to assortative.
\newblock {\em Phys. Rev. E}, 70:066102, Dec 2004.

\bibitem{Horn2012}
Roger~A Horn and Charles~R Johnson.
\newblock {\em Matrix analysis}.
\newblock Cambridge university press, 2012.

\bibitem{Karrer2010}
Brian Karrer and M.~E.~J. Newman.
\newblock Message passing approach for general epidemic models.
\newblock {\em Phys. Rev. E}, 82:016101, Jul 2010.

\bibitem{Lokhov2014}
Andrey~Y. Lokhov, Marc M\'ezard, Hiroki Ohta, and Lenka Zdeborov\'a.
\newblock Inferring the origin of an epidemic with a dynamic message-passing
  algorithm.
\newblock {\em Phys. Rev. E}, 90:012801, Jul 2014.

\bibitem{Ryan2015}
Ryan~A. Rossi and Nesreen~K. Ahmed.
\newblock The network data repository with interactive graph analytics and
  visualization.
\newblock In {\em AAAI}, 2015.

\bibitem{Cios2005}
Krzysztof~J Cios and Lukasz~A Kurgan.
\newblock Trends in data mining and knowledge discovery.
\newblock In {\em Advanced techniques in knowledge discovery and data mining},
  pages 1--26. Springer, 2005.

\bibitem{Batagelj2006}
Vladimir Batagelj and Andrej Mrvar.
\newblock Analysis of large networks.
\newblock In {\em Proceedings of Pajek Workshop at XXVI Sunbelt Conference},
  2006.

\bibitem{Leskovec2014}
Jure Leskovec and Andrej Krevl.
\newblock {SNAP Datasets}: {Stanford} large network dataset collection.
\newblock \url{http://snap.stanford.edu/data}, June 2014.

\bibitem{Barabasi2016}
Albert-L{\'a}szl{\'o} Barab{\'a}si.
\newblock {\em Network Science}.
\newblock Cambridge University Press, 2016.

\bibitem{Goh2001}
K.-I. Goh, B.~Kahng, and D.~Kim.
\newblock Universal behavior of load distribution in scale-free networks.
\newblock {\em Phys. Rev. Lett.}, 87:278701, Dec 2001.

\end{thebibliography}
\end{document}